%% file: ms.tex
\documentclass[12pt,preprint2]{aastex}

\shorttitle{Off--Axis Bulge Abundances}
\shortauthors{Johnson et al.}

\begin{document}

\title{METALLICITY DISTRIBUTION FUNCTIONS, RADIAL VELOCITIES, AND ALPHA
ELEMENT ABUNDANCES IN THREE OFF--AXIS BULGE FIELDS}

\author{
Christian I. Johnson\altaffilmark{1,6,7,8},
R. Michael Rich\altaffilmark{1},
Chiaki Kobayashi\altaffilmark{2},
Andrea Kunder\altaffilmark{3},
Catherine A. Pilachowski\altaffilmark{4},
Andreas Koch\altaffilmark{5}, and
Roberto de Propris\altaffilmark{3}
}

\altaffiltext{1}{Department of Physics and Astronomy, UCLA, 430 Portola Plaza,
Box 951547, Los Angeles, CA 90095-1547, USA; cijohnson@astro.ucla.edu;
rmr@astro.ucla.edu}

\altaffiltext{2}{Centre for Astrophysics Research, University of Hertfordshire,
Hatfield  AL10 9AB, UK; c.kobayashi@herts.ac.uk}

\altaffiltext{3}{Cerro Tololo Inter--American Observatory, Casilla 603, La 
Serena, Chile; akunder@ctio.noao.edu}

\altaffiltext{4}{Department of Astronomy, Indiana University, Swain West 319,
727 East Third Street, Bloomington, IN 47405-7105, USA;
catyp@astro.indiana.edu}

\altaffiltext{5}{Zentrum f\"{u}r Astronomie der Universit\"{a}t 
Heidelberg, Landessternwarte, K\"{o}nigstuhl 12, Heidelberg, Germany;
akoch@lsw.uni-heidelberg.de}

\altaffiltext{6}{Visiting astronomer, Cerro Tololo Inter--American
Observatory, National Optical Astronomy Observatory, which are operated by the
Association of Universities for Research in Astronomy, under contract with the
National Science Foundation.}

\altaffiltext{7}{Visiting Astronomer, Kitt Peak National Observatory, National 
Optical Astronomy Observatories, which is operated by the Association of 
Universities for Research in Astronomy, Inc. (AURA) under cooperative agreement
with the National Science Foundation.  The WIYN Observatory is a joint facility
of the University of Wisconsin--Madison, Indiana University, Yale University, 
and the National Optical Astronomy Observatory.}

\altaffiltext{8}{National Science Foundation Astronomy and Astrophysics
Postdoctoral Fellow}

\begin{abstract}

We present radial velocities and chemical abundance ratios of [Fe/H], [O/Fe], 
[Si/Fe], and [Ca/Fe] for 264 red giant branch (RGB) stars in three Galactic 
bulge off--axis fields located near (l,b)=(--5.5,--7), (--4,--9), and 
($+$8.5,$+$9).  The results are based on equivalent width and spectrum
synthesis analyses of moderate resolution (R$\approx$18,000), high 
signal--to--noise ratio (S/N$\sim$75--300 pixel$^{\rm -1}$) spectra obtained 
with the Hydra spectrographs on the Blanco 4m and WIYN 3.5m telescopes.  The 
targets were selected from the blue side of the giant branch to avoid cool
stars that would be strongly affected by CN and TiO; however, a comparison of
the color--metallicity distribution in literature samples suggests our 
selection of bluer targets should not present a significant bias against 
metal--rich stars.  We find a full range in metallicity that spans 
[Fe/H]$\approx$--1.5 to $+$0.5, and that, in accordance with the previously 
observed minor--axis vertical metallicity gradient, the median [Fe/H] also 
declines with increasing Galactic latitude in off--axis fields.  The off--axis 
vertical [Fe/H] gradient in the southern bulge is estimated to be $\sim$0.4 dex
kpc$^{\rm -1}$; however, comparison with the minor--axis data suggests a strong
radial gradient does not exist.  The ($+$8.5,$+$9) field exhibits a higher than
expected metallicity, with a median [Fe/H]=--0.23, that might be related to a 
stronger presence of the X--shaped bulge structure along that line--of--sight. 
This could also be the cause of an anomalous increase in the median radial 
velocity for intermediate metallicity stars in the ($+$8.5,$+$9) field.  
However, the overall radial velocity and dispersion for each field are in good 
agreement with recent surveys and bulge models.  All fields exhibit an 
identical, strong decrease in velocity dispersion with increasing metallicity 
that is consistent with observations in similar minor--axis outer bulge fields.
Additionally, the [O/Fe], [Si/Fe], and [Ca/Fe] versus [Fe/H] trends are 
identical among our three fields, and are in good agreement with past bulge 
studies.  We find that stars with [Fe/H]$\la$--0.5 are $\alpha$--enhanced, and 
that the [$\alpha$/Fe] ratios decline at higher metallicity.  At [Fe/H]$\la$0, 
the $\alpha$--element trends are indistinguishable from the halo and thick 
disk, and the variations in the behavior of individual $\alpha$--elements are 
consistent with production in massive stars and a rapid bulge formation 
timescale.

\end{abstract}


\section{INTRODUCTION}

Although the minor--axis of the Galactic bulge has been the subject of 
extensive high resolution spectroscopic study (e.g., Zoccali et al. 2008), 
there has been relatively little chemical characterization of stars in 
off--axis bulge fields.  Iron and $\alpha$--element abundances are reported by 
Gonzalez et al. (2011) in a field near the globular cluster NGC 6553 
(l,b)=($+$5.25,--3), but that is the only off--axis field for which 
significant (N$\ga$100 stars) chemical abundance data are presently 
available\footnote{Some results from the ARGOS study are presented in Ness 
et al. (2013), but the individual abundances are not yet publicly available.}. 
Overall, the radial variation in [Fe/H]\footnote{We make use of the standard 
spectroscopic notation where 
[A/B]$\equiv$log(N$_{\rm A}$/N$_{\rm B}$)$_{\rm star}$--log(N$_{\rm A}$/N$_{\rm B}$)$_{\sun}$ and log $\epsilon$(A)$\equiv$log(N$_{\rm A}$/N$_{\rm H}$)+12.0
for elements A and B.} and [$\alpha$/Fe] along the bulge major--axis and 
off--axis is poorly constrained.

It is also clear that the bulge is spatially complex, and a deep understanding 
will require a full three--dimensional sampling of the structure.  McWilliam \& 
Zoccali (2010) found red clump giants to be spatially distributed in an 
X--shaped structure that spans the inner $\sim$8$\degr$, and was confirmed in 
a density analysis by Saito et al. (2011).  The metal--rich subset of the 
bulge appears to be predominant in the X--shaped structure (Ness et al. 2012;
Uttenthaler et al. 2012; but see also de Propris et al. 2011).

Connected with these issues is the question of whether there is kinematic 
substructure in the bulge, and whether the population is comprised of 
multiple components.  Some evidence suggests the bulge may host two dominant
populations (Babusiaux et al. 2010; Bensby et al. 2011; Hill et al. 2011; 
Uttenthaler et al. 2012).  At the most basic level, such a bimodality might be 
expressed in the bar and the ``classical" bulge.  Soto et al. (2007) argued 
that the metal--rich bulge stars exhibit a vertex deviation expected for stars 
orbiting in the bar, while the metal--poor ([Fe/H]$<$--0.5) stars could 
plausibly be on non--barred orbits.  Hill et al. (2011) argues strongly for 
bimodality, connecting two putative peaks in the [Fe/H] distribution with 
differing [Mg/Fe] ratios.  Bensby et al. (2011) also find evidence for 
bimodality in their sample of 26 microlensed bulge dwarfs and subgiants, 
although they have recently increased their sample to 58 stars (Bensby et al.
2013) and now favor a multi--component model that may also incorporate the 
bimodal structures found in previous studies.  Similarly, Ness et al. (2013), 
using [Fe/H] values derived from moderate resolution spectra of a large sample 
($\sim$28,000) of bulge giant and clump stars, deconvolve the metallicity 
distributions into 5 separate Gaussians that are asserted to vary 
systematically with Galactic latitude and longitude, and to reflect different 
populations.  The BRAVA survey of $\sim$10,000 bulge M giants (e.g., Rich et 
al. 2007a; Kunder et al. 2012) does not find any hint of kinematic substructure
or a significant classical bulge.  Shen et al. (2010) argue that very strong 
limits are placed on any classical bulge fraction ($\la$8--15$\%$ of the disk
mass) from the BRAVA survey.  It has been shown theoretically possible for a 
rapidly rotating bar to spin up a classical bulge (e.g., Shen et al. 2010; 
Saha et al. 2012), but such an effect would likely be unveiled by the 
metal--poor stars exhibiting rapid rotation and bar--like kinematics, which is 
not observed.

The question of correlations between [Fe/H], velocity, and/or velocity 
dispersion are among the most actively debated aspects of the substructure 
argument.  Babusiaux et al. (2010) find a striking trend for the more 
metal--rich stars to have the highest velocity dispersion in Baade's Window
(b=--4$\degr$); the trend reverses itself by b=--12$\degr$.  Exploring 
intermediate minor--axis fields near b=--8$\degr$ and --10$\degr$, Johnson et 
al. (2011) and Uttenthaler et al. (2012) find the common trend (e.g. Rich 
1990) of velocity dispersion decreasing with increasing metallicity.

A further motivation to explore off--axis bulge fields addresses similarities 
and differences between such fields and the Galactic thick disk.  In a bulge 
formation scenario in which the bar has thickened from a preexisting massive 
disk, it would not be surprising to find chemical similarities between bulge 
and thick disk stars; these might even be expected to predominate off--axis.
Here, there is also some uncertainty.  While early analyses (Zoccali et 
al. 2006; Fulbright et al. 2007; Lecureur et al. 2007) found that bulge stars
were likely $\alpha$--enhanced, particularly in their [O/Fe] and [Mg/Fe] 
ratios, compared to the thick disk, more recent differential studies (e.g., 
Mel{\'e}ndez et al. 2008; Ryde et al. 2010; Alves--Brito et al. 2010; Gonzalez 
et al. 2011) have instead concluded that the thick disk and metal--poor bulge 
likely share identical [$\alpha$/Fe] distributions (but see also Bensby et al.
2013).  However, it is not yet clear if these abundance similarities extend to 
super--solar metallicities and/or light odd--Z and heavy neutron--capture 
elements (e.g., Fulbright et al. 2007; Lecureur et al. 2007; Alves--Brito et 
al. 2010; McWilliam et al. 2010; Johnson et al. 2012).

Exploring the bulge in new fields, and near the bulge/halo boundary, offers 
significant opportunities to search for bulge substructure and to explore 
whether the bulge population is complex and multimodal over its whole volume.  
Although future samples will reach thousands of stars, our sample of 264 red
giant branch (RGB) stars observed in multiobject mode at moderate 
(R$\sim$18,000) resolution will yield new constraints on the structure and 
substructure of the bulge.  These data will also provide new tests of whether 
the bulge resembles the thick disk, and the presence of kinematic or 
composition substructure.

\section{OBSERVATIONS AND DATA REDUCTION}

The observations for this project were obtained between 2011 August 19--20 with
the WIYN 3.5m telescope at Kitt Peak National Observatory and 2011 September 
8--12 with the Blanco 4m telescope at Cerro Tololo Inter--American Observatory.
WIYN was used to observe the northern bulge field centered near 
(l,b)=($+$8.5,$+$9) and the Blanco was used to observe the two southern bulge 
fields centered near (l,b)=(--5.5,--7) and (--4,--9).  All spectra were 
acquired using the Hydra multifiber positioners and bench spectrographs.  For 
WIYN--Hydra, we used the red fibers, 316 line mm$^{\rm -1}$ Echelle grating, 
red camera, and X18 filter to achieve a resolving power of 
R($\lambda$/$\Delta$$\lambda$)$\approx$18,000.  We also employed only one 
spectrograph setup with wavelength coverage ranging from about 6050--6350 \AA.
For Blanco--Hydra, we used the large 300$\micron$ (2$\arcsec$) fibers, 400 mm
Bench Schmidt camera, 316 line mm$^{\rm -1}$ Echelle grating, and the 
E6257 and E6757 filters to achieve R$\approx$18,000 in the first setup 
(6125--6345 \AA) and R$\approx$15,000 in the second setup (6550--6840 \AA).
A summary of the observation dates, instrument setups, and exposure times for
each field is provided in Table 1.

The data reduction for both WIYN--Hydra and Blanco--Hydra data were carried out
using standard IRAF tasks.\footnote{IRAF is distributed by the National Optical
Astronomy Observatory, which is operated by the Association of Universities for
Research in Astronomy, Inc., under cooperative agreement with the National 
Science Foundation.}  Overscan removal and bias subtraction were accomplished
using the \emph{ccdproc} routine.  The remaining data reduction tasks, which
include fiber identification and tracing, scattered light removal, flat--field
correction, ThAr wavelength calibration, cosmic--ray removal, sky subtraction,
and object spectrum extraction, were carried out using the \emph{dohydra}
task.  For the spectrograph setups that included the $\sim$6275--6330 \AA\ 
telluric absorption feature, we observed several rapidly rotating B stars at
various airmasses in order to remove telluric contamination in the object 
spectra.  The final telluric correction was applied with the \emph{telluric}
IRAF task.  All spectra were then combined via co--addition and continuum 
normalized by fitting a low order polynomial through the continuum points 
identified in Fulbright et al. (2006; their Table 5).  The signal--to--noise 
ratio (S/N) of the combined spectra ranged from 75 to more than 300.

\subsection{Target Selection}

Since many of the previous bulge spectroscopic analyses have focused primarily
on minor--axis fields (see $\S$1), we targeted fields that were located away 
from the minor--axis (i.e., l$\neq$0).  The observed fields were also selected
to have relatively low interstellar extinction (E(B--V)$\la$0.35), in order to
minimize the integration time needed to reach S/N$\ga$75.  For the Blanco, 5 
hours of integration with our Hydra setups yields S/N$\ga$75 for stars with 
V$\la$14.5.  Given the higher airmass and shorter time frame with which we were
able to observe the bulge with WIYN, we selected brighter targets 
(V$\la$13.5--14).  However, the significant throughput increase provided by the
recent upgrade to the WIYN bench spectrograph (Bershady et al. 2008; Knezek et 
al. 2010) proved to make the selection of brighter stars an unnecessary 
constraint.

Unfortunately, constraining bulge membership for individual stars is often a
difficult task using currently available data.  Although high--quality 
optical photometry is not available for large regions of the bulge, the
2MASS database (Skrutskie et al. 2006) provides accurate coordinates and 
infrared photometry for the bright giants we were targeting.  From the raw 
2MASS catalog, we created an initial target list by only selecting stars which 
had: (1) 4000$<$T$_{\rm eff}$$<$4800 K, (2) V$<$14.5, and (3) log(g)$<$2.5.
The temperatures were determined using the J--K color--temperature
relation provided by Alonso et al. (1999; 2001).  All stars were individually
dereddened using the E(B--V) values derived from the Schlegel et al. (1998)
dust maps.  Optical V magnitudes were estimated by taking the T$_{\rm eff}$
values calculated from J--K and then inverting the Alonso et al. (1999) V--K 
color--temperature relation to solve for V.  The range of possible log(g) 
values for each star were determined by examining the 
temperature--gravity--color relations in Ku{\v c}inskas et al. (2005; 2006)
to ensure log(g)$<$2.5.\footnote{Since the temperature--gravity--color 
relations in Ku{\v c}inskas et al. (2005; 2006) are only calibrated up to
[Fe/H]=0, we applied a linear extrapolation out to [Fe/H]=$+$0.5.  This permits
coverage from [Fe/H]=--2 to $+$0.5, which spans the bulge's full metallicity
range.}  The culled preliminary target list of $\sim$1000--3000 stars was then
input into the Hydra fiber positioning software to produce an optimal 
configuration file for each field.  In addition to $\sim$10--20 sky fibers for
each configuration, we were able to place fibers on 91 stars for the 
(--5.5,--7) field, 105 stars for the (--4,--9) field, and 68 stars for the 
($+$8.5,$+$9) field.

\begin{figure}
\epsscale{1.00}
\plotone{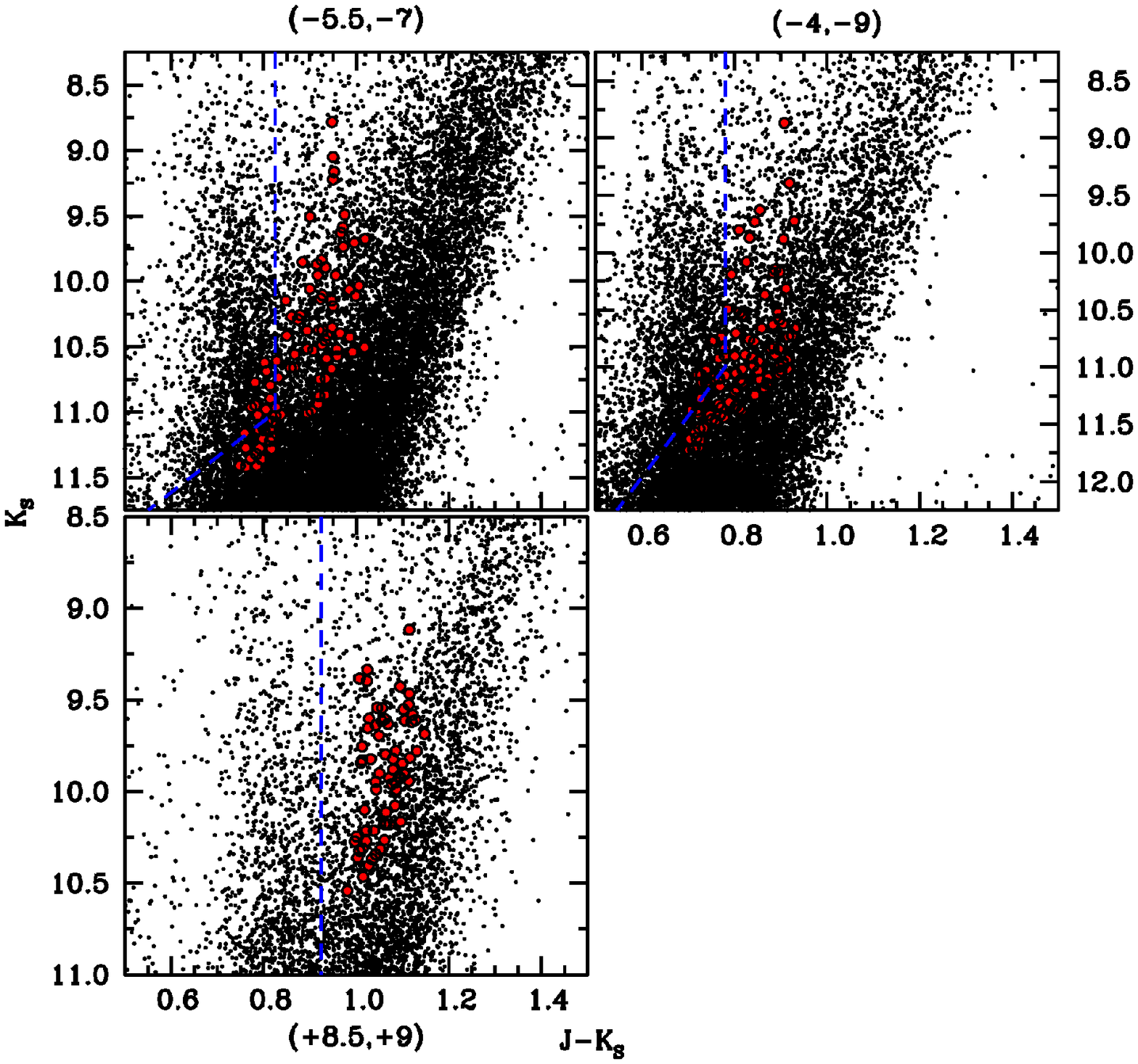}
\caption{K$_{\rm S}$ versus J--K$_{\rm S}$ color--magnitude diagrams for the
three observed bulge fields.  The small black circles are point sources from
the 2MASS catalog, and the filled red circles are the stars observed in the 
(l,b)=(--5.5$\degr$,--7$\degr$), (--4$\degr$,--9$\degr$), and
($+$8.5$\degr$,$+$9$\degr$) fields.  The dashed blue lines designate the 
approximate separation between the bulge RGB and the foreground red clump 
population.}
\label{f1}
\end{figure}

In Figure \ref{f1} we show 2MASS color--magnitude diagrams (CMDs) for all 
three fields, and also identify the observed target stars.  While $\sim$90$\%$ 
of the targets in the two southern bulge fields and all of the stars in the 
northern bulge field lie well on the bulge RGB, $\sim$10$\%$ of the stars in 
the two southern bulge fields lie in a region that is likely a mix of bulge 
RGB and foreground red clump stars (the approximate separation is illustrated 
by the dashed blue lines in Figure \ref{f1}).  It is possible that some of 
these targets are actually foreground red clump stars, but we find only a 
small fraction of the bluest stars to have large proper motions and estimate 
the contamination ratio to be $\la$10$\%$ of the total sample (see $\S$4.1.1). 
For the majority of target stars that lie on the RGB, it is clear from Figure 
\ref{f1} that the observations are biased toward the bluer side of the RGB.  
This is a result of our choice to not select stars with T$_{\rm eff}$$<$4000 K,
in order to avoid strong CN and TiO molecular features in the spectra that 
would prevent reliable abundance determinations via equivalent width 
measurements.  However, we show in $\S$4.1.1 that we do not expect the 
selection of bluer stars to produce a strong metallicity bias.

\section{DATA ANALYSIS}

\subsection{Model Stellar Atmospheres}

As mentioned above, we used the dereddened (J--K$_{\rm S}$)$_{\rm o}$ 
color\footnote{The 2MASS J and K$_{\rm S}$ values were converted onto the 
TCS system, as required for using the Alonso et al. (1999) color--temperature
relation.} for each star to derive T$_{\rm eff}$ via the Alonso et al. (1999)
color--temperature relation.  In order to account for interstellar extinction,
we assumed the relation E(J--K$_{\rm S}$)/E(B--V)=0.505 (Fiorucci \& Munari 
2003) and used the recommended E(B--V) value from Schlegel et al. 
(1998).\footnote{The E(B--V) values can be accessed at: 
http://irsa.ipac.caltech.edu/applications/DUST/.}  The average E(B--V) values
were 0.240 ($\sigma$=0.036), 0.136 ($\sigma$=0.008), and 0.385 ($\sigma$=0.024)
for the (--5.5,--7), (--4,--9), and ($+$8.5,$+$9) fields, respectively.  We did
not attempt to refine T$_{\rm eff}$ using excitation equilibrium; however, a 
star--by--star examination of the log $\epsilon$(Fe I) abundance versus 
excitation potential plot indicated that significant ($\ga$100 K) T$_{\rm eff}$
adjustments were likely not necessary.  This was true even for the most 
metal--rich stars ([Fe/H]$>$$+$0.2), which fall just outside the calibration 
range of the Alonso et al. (1999) J--K color--temperature relation.  In a 
more quantitative sense, the average, median, and standard deviations of the
slopes for the log $\epsilon$(Fe I) versus excitation potential plots were:
--0.006, --0.011, and 0.043 for the (--5.5,--7) field, $+$0.002, --0.004, and
0.032 for the (--4,--9) field, and --0.019, --0.023, and 0.037 for the 
($+$8.5,$+$9) field.  For comparison, altering T$_{\rm eff}$$\pm$100 K in 
a ``typical" bulge giant with [Fe/H]=--0.3 is expected to produce a change in 
the magnitude of the slope by $\sim$0.04 at T$_{\rm eff}$=4200 K and 
$\sim$0.02 at T$_{\rm eff}$=4800 K.

\begin{figure}
\epsscale{1.00}
\plotone{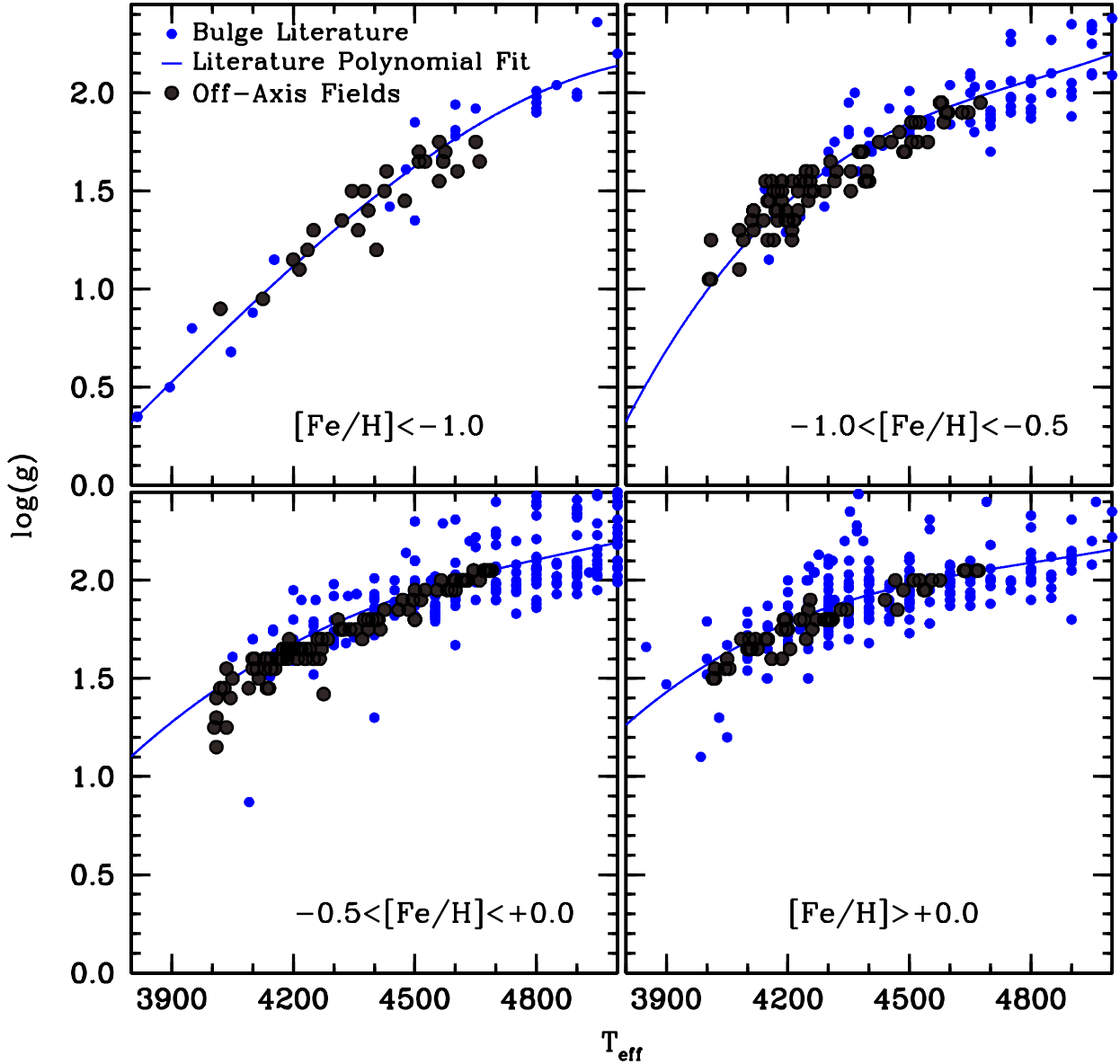}
\caption{The four panels show the relationship between log(g) 
and T$_{\rm eff}$ for bulge RGB stars in our combined sample (filled grey 
circles) and those available in the literature (filled blue circles).  Each
panel contains only stars within the listed metallicity range.
A 3$^{\rm rd}$ order polynomial is fit to the literature data and is 
illustrated by the solid blue lines.  The literature data are from: McWilliam
\& Rich (1994), Rich \& Origlia (2005), Fulbright et al. (2006), Lecureur et 
al. (2007), Rich et al. (2007b), Mel{\'e}ndez et al. (2008), Zoccali et al. 
(2008), Alves--Brito et al. (2010), Ryde et al. (2010), Johnson et al. (2011), 
and Rich et al. (2012).}
\label{f2}
\end{figure}

Surface gravities were derived by interpolating within the 11 Gyr isochrones 
provided by the Dartmouth Stellar Evolution Database (Dotter et al. 
2007).\footnote{The isochrones can be accessed at: 
http://stellar.dartmouth.edu/~models/grid.html.}  We initially assumed a level 
of $\alpha$--enhancement that followed the [$\alpha$/Fe] versus [Fe/H] trends 
of Gonzalez et al. (2011) and Johnson et al. (2011).  However, after measuring 
[$\alpha$/Fe] for each star we reinterpolated using a grid with the appropriate
level of $\alpha$--enhancement.  As can be seen in Figure \ref{f2}, our assumed
log(g) values are typically in good agreement with other log(g) estimates of 
bulge RGB stars in the literature.  Note that the literature values shown in 
Figure \ref{f2} represent a variety of techniques for deriving log(g), 
including ionization equilibrium, isochrone fitting, and photometric estimates.
%
%
%
%

For each star the model atmosphere metallicity was set at [M/H]=--0.3 and then
adjusted to equal the average derived [Fe/H] ratio.  Similarly, the 
microturbulence (vt) was set at 2 km s$^{\rm -1}$ and then adjusted so that 
the plot of Fe I abundance versus line strength produced a zero slope.  For 
all stars the model atmosphere was calculated by interpolating within the 
$\alpha$--rich AODFNEW ATLAS9 grid (Castelli et al. 1997).\footnote{The model
atmosphere grid can be accessed at: 
http://wwwuser.oat.ts.astro.it/castelli/grids.html.}  While not all of the 
stars in our sample are $\alpha$--rich, Fulbright et al. (2007) found that
the impact of using the solar scaled versus $\alpha$--rich ATLAS9 models on 
derived abundances was relatively small for bulge giants ($\la$0.1 dex in most 
cases).  Specifically, the average effect on [Fe/H], [O/Fe], [Si/Fe], and 
[Ca/Fe] was found to be (in the sense AODFNEW--ODFNEW): $+$0.06 
($\sigma$=0.03), --0.05 ($\sigma$=0.07), $+$0.02 ($\sigma$=0.03), and 
--0.02 ($\sigma$=0.03), respectively.  We performed the same test on our data
set and found that the average AODFNEW--ODFNEW abundance change for [Fe/H], 
[O/Fe], [Si/Fe], and [Ca/Fe] was similar at: $+$0.05 ($\sigma$=0.07), $+$0.06 
($\sigma$=0.06), $+$0.10 ($\sigma$=0.03), and --0.10 ($\sigma$=0.03), 
respectively.  The final model atmosphere parameters, photometry, and E(B--V)
values are provided in Table 2.

\subsection{Equivalent Width Abundance Determinations}

The Fe I abundances were determined by measuring equivalent widths (EWs), using
software developed for Johnson et al. (2008).  The final [Fe/H] abundances, 
listed in Tables 2 and 3, were derived using the 2010 version of the LTE 
line analysis code MOOG (Sneden 1973).  Single, isolated lines were 
fit with a Gaussian profile and moderately blended lines were deblended with
up to five Gaussian profiles.  We selected 61 Fe I lines located in the 
wavelength regions specified in $\S$2 that were relatively isolated in the 
spectra of both Arcturus (Hinkle et al. 2000)\footnote{The Arcturus atlas can 
be accessed at: http://www.noao.edu/archives.html.} and $\mu$ Leo (Moultaka et 
al. 2004).\footnote{Based on spectral data retrieved from the 
ELODIE archive at Observatoire de Haute-Provence (OHP).  The ELODIE archive
can be accessed at: http://atlas.obs-hp.fr/elodie/.}  On average, the [Fe/H]
abundances for the (--5.5,--7), (--4,--9), and ($+$8.5,$+$9) fields are based
on 21 Fe I lines.  The number of lines used for each star is less than the 61
potential lines due to variations in wavelength coverage (Blanco--Hydra versus
WIYN--Hydra), temperature, metallicity, and S/N.  The average line--to--line 
dispersion of 0.16 dex ($\sigma$=0.03) was found to not vary significantly
as a function of temperature or metallicity, and is reasonable given the 
relatively high metallicity and cool temperatures of our target stars coupled 
with the moderate spectral resolution.

\begin{figure}
\epsscale{0.75}
\plotone{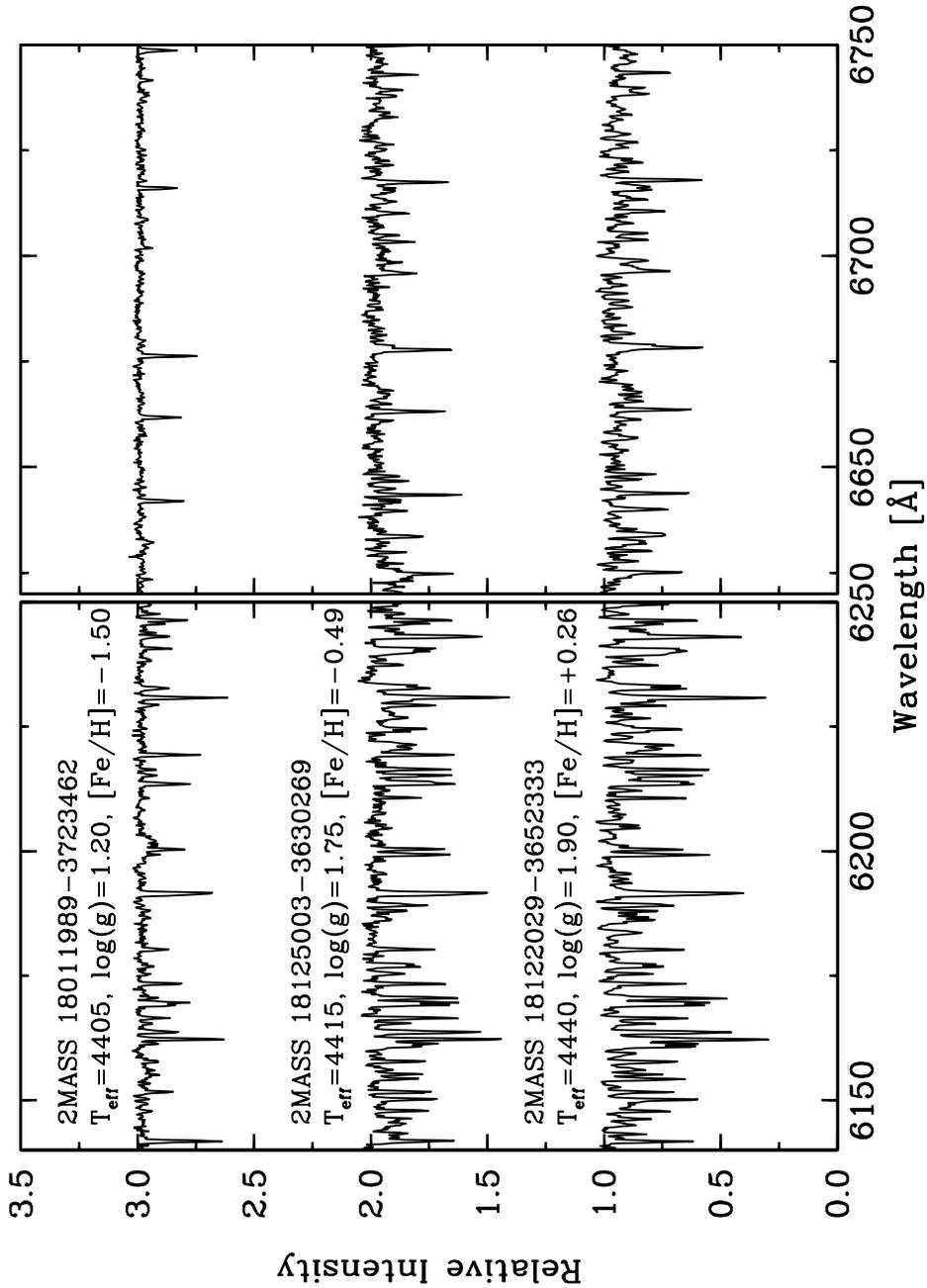}
\caption{Typical sample spectra for stars of comparable T$_{\rm eff}$ are 
shown to illustrate how the continuum windows, line strengths, and molecular 
features change as a function of metallicity.  The left panel shows a portion
of the 6300 \AA\ region spectra, and the right panel shows a portion of the 
6700 \AA\ region spectra for the same three stars.  The spectra have been 
offset for display purposes.}
\label{f3}
\end{figure}

The Fe I abundances were determined on a line--by--line basis relative to 
Arcturus.  The oscillator strength (log gf) values listed in Table 4 were 
derived by measuring the EW of each line in the high resolution, high S/N 
Arcturus atlas and forcing the derived abundance for each line to be 
[Fe/H]=--0.50, assuming the Arcturus model atmosphere parameters given in 
Fulbright et al. (2006; T$_{\rm eff}$=4290 K, log(g)=1.60, [Fe/H]=--0.50, and 
vt=1.67 km s$^{\rm -1}$).  The wavelength and excitation potential values given
in Table 4 are from the NIST Atomic Spectra Database (Ralchenko et al. 
2011)\footnote{The NIST database can be 
accessed at: http://www.nist.gov/pml/data/asd.cfm.}, Vienna Atomic Line 
Database (VALD; Kupka et al. 2000)\footnote{VALD can be accessed at: 
http://vald.astro.univie.ac.at/~vald/php/vald.php.}, or Thevenin (1990).
Sample spectra of both the 6250 and 6700 \AA\ setups for stars with similar 
T$_{\rm eff}$ but different [Fe/H] are shown in Figure \ref{f3}.  This figure
also illustrates data quality as well as the change in line strength, 
continuum availability, and molecular contamination as a function of [Fe/H].

\subsection{Spectrum Synthesis Abundance Determinations}

For the elements other than Fe, we determined abundances via spectrum synthesis
rather than EW measurement.  Given the large number of stars analyzed here, 
we employed the computationally parallel version of the MOOG \emph{synth} 
driver developed for Johnson et al. (2012).  Similar to the Fe abundance
measurements, [O/Fe], [Si/Fe], and [Ca/Fe] ratios were determined relative to
Arcturus.  We adopted the Arcturus abundances derived in Fulbright et al. 
(2007): [O/Fe]=$+$0.48, [Si/Fe]=$+$0.35, and [Ca/Fe]=$+$0.21.  These adopted 
values are within $\sim$0.1 dex of those derived recently by Ram{\'{\i}}rez \& 
Allende Prieto (2011).  The lines used here are listed in Table 3.

\begin{figure}
\epsscale{0.75}
\plotone{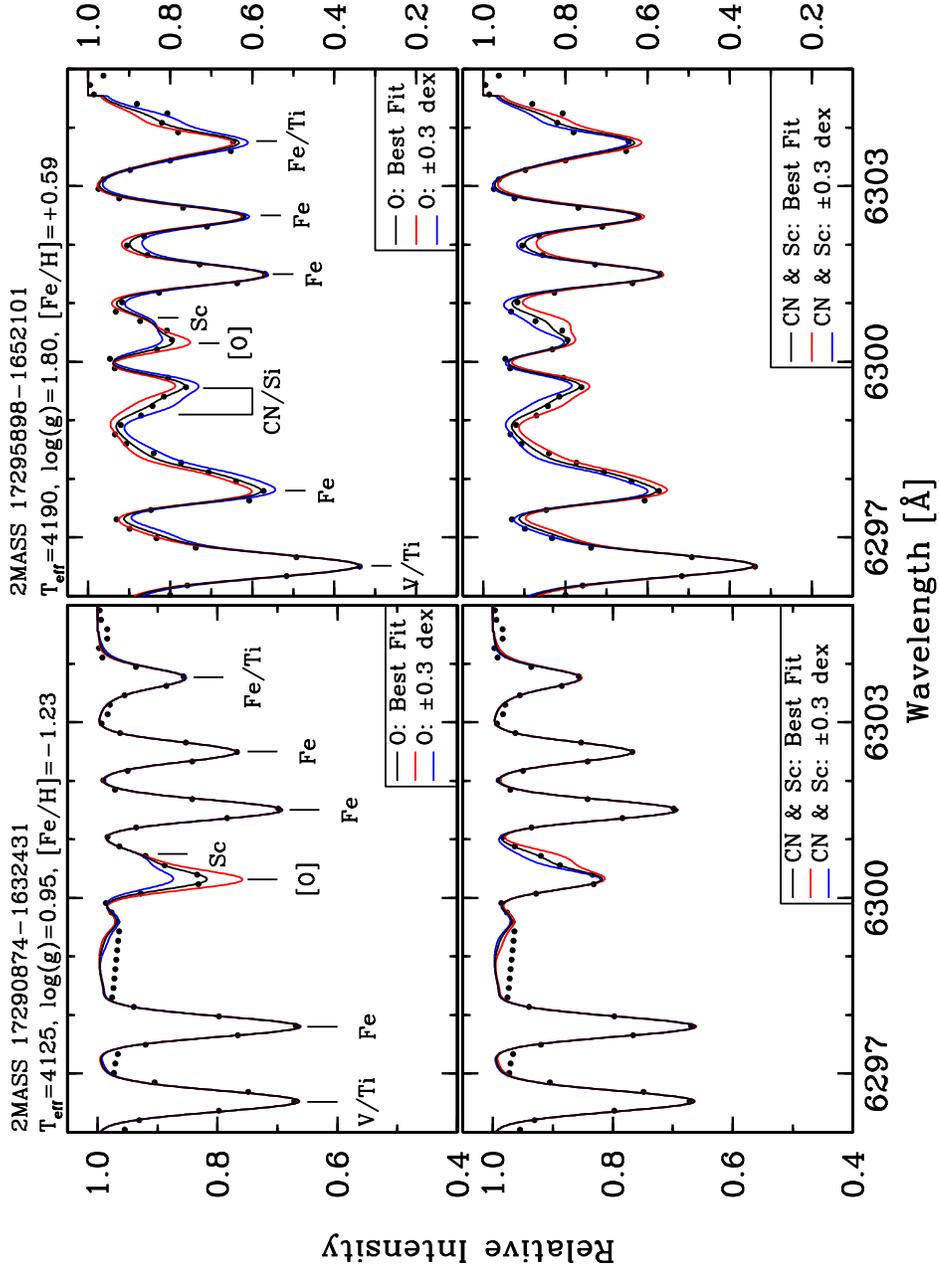}
\caption{Sample spectrum synthesis fits to the region around the 6300\AA\ [O I]
line for a cool, metal--poor star (left panels) and a cool metal--rich star
(right panels).  In the top left and top right panels, the solid black line
shows the best fit oxygen abundance.  The colored solid lines illustrate
how the syntheses change when the oxygen abundance is altered by $+$0.3 dex
(red lines) and --0.3 dex (blue lines) from the best--fit value.  Similarly,
the bottom left and bottom right panels show how the syntheses change when
oxygen is held fixed at the best--fit value while the CN and Sc abundances
are altered by $\pm$0.3 dex.}
\label{f4}
\end{figure}

For the 6300 \AA\ [O I] line we synthesized the region spanning 6295--6305 \AA,
including the 6300.34 \AA\ Ni I line that is blended with the 6300.30 [O I]
feature.  We assumed [Ni/Fe]=0 for all syntheses.  The initial line list was 
compiled from VALD for atomic lines and the Kurucz database\footnote{The Kurucz
line list database can be accessed at: 
http://kurucz.harvard.edu/linelists.html.} for CN molecular lines.  The 
final log gf values for all significant atomic and molecular lines in the 
6295--6305 \AA\ region were adjusted to minimize the difference between our
synthetic spectrum and the Arcturus atlas.  In order to fit the CN features,
we adopted the Arcturus carbon and nitrogen abundances from Peterson et al. 
(1993; [C/Fe]=$+$0.0 and [N/Fe]=$+$0.3).  In general this provided a 
satisfactory fit to the CN lines using the  original Kurucz line list and only 
a few minor adjustments were required.  For the Sc I line that is blended 
with the [O I] feature in our Hydra spectra but not the Arcturus atlas, we set
the log gf value assuming [Sc/Fe]=$+$0.2 (Peterson et al. 1993).  However, as
can be seen in the sample syntheses provided in Figure \ref{f4}, there is 
enough separation between the [O I] and Sc features in the Hydra spectra to 
reasonably account for the Sc contribution.  

\begin{figure}
\epsscale{0.75}
\plotone{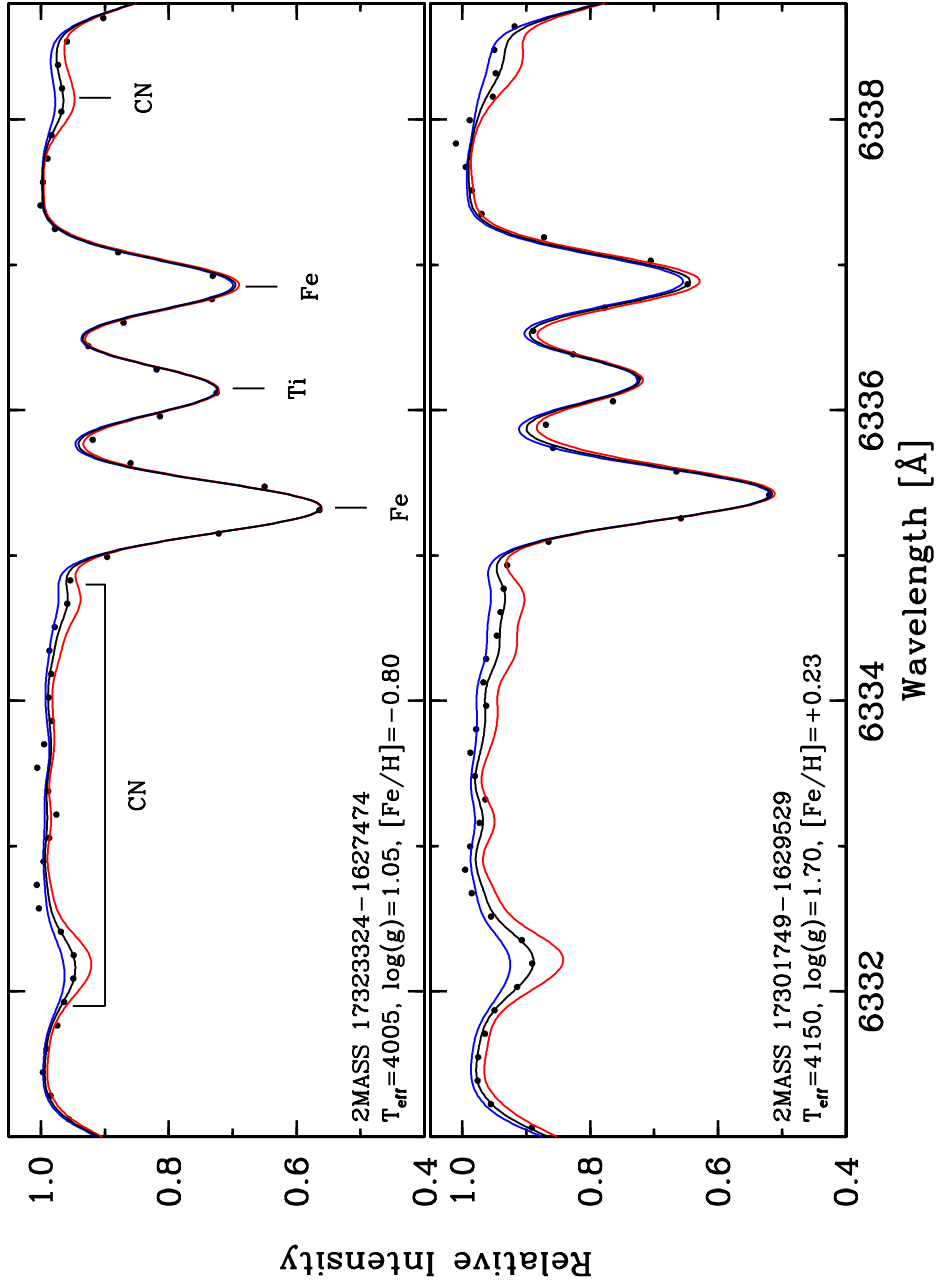}
\caption{Similar to Figure \ref{f4}, the synthetic spectrum fits shown here
illustrate how the CN abundance was determined for each star.  The top panel
shows a cool, metal--poor example and the bottom panel shows a cool, 
metal--rich example.  The black line indicates the best fit CN abundance.
The red and blue syntheses indicate changes in the CN abundance by $\pm$0.3 
dex, respectively.  Note that except for a weak Si I and Fe II blend near
6332 \AA, the two regions spanning 6331.5--6335 \AA\ and 6337.5--6339.8 \AA\
contain almost exclusively CN features.}
\label{f5}
\end{figure}

Since the [O I] line is both modestly affected by CN blending (see bottom 
panels of Figure \ref{f4}) and also by the C$+$N abundance (via 
the molecular equilibrium calculation), we identified a small window in the 
6250 \AA\ spectra that could be used to roughly estimate the CN abundance.
While there are other similar windows that may be used, especially in the 
6700 \AA\ spectra, we selected the 6331--6339 \AA\ window (shown in 
Figure \ref{f5}) because it provided a several Angstrom wide region comprised
of mostly CN lines and was observed with both the WIYN and Blanco bench 
spectrographs.  For each star we initially set [C/Fe]=--0.30 
and [N/Fe]=$+$0.50, values typical for bulge RGB stars (e.g., Mel{\'e}ndez et 
al. 2008; Ryde et al. 2010), and then adjusted [N/Fe] until a satisfactory fit 
was found.  The derived [C/Fe] and [N/Fe] abundances were then used in the 
[O I] synthesis.  This processes was iterated at least once for each star to 
account for the correlated variations in line strength between CN and O.

\begin{figure}
\epsscale{0.90}
\plotone{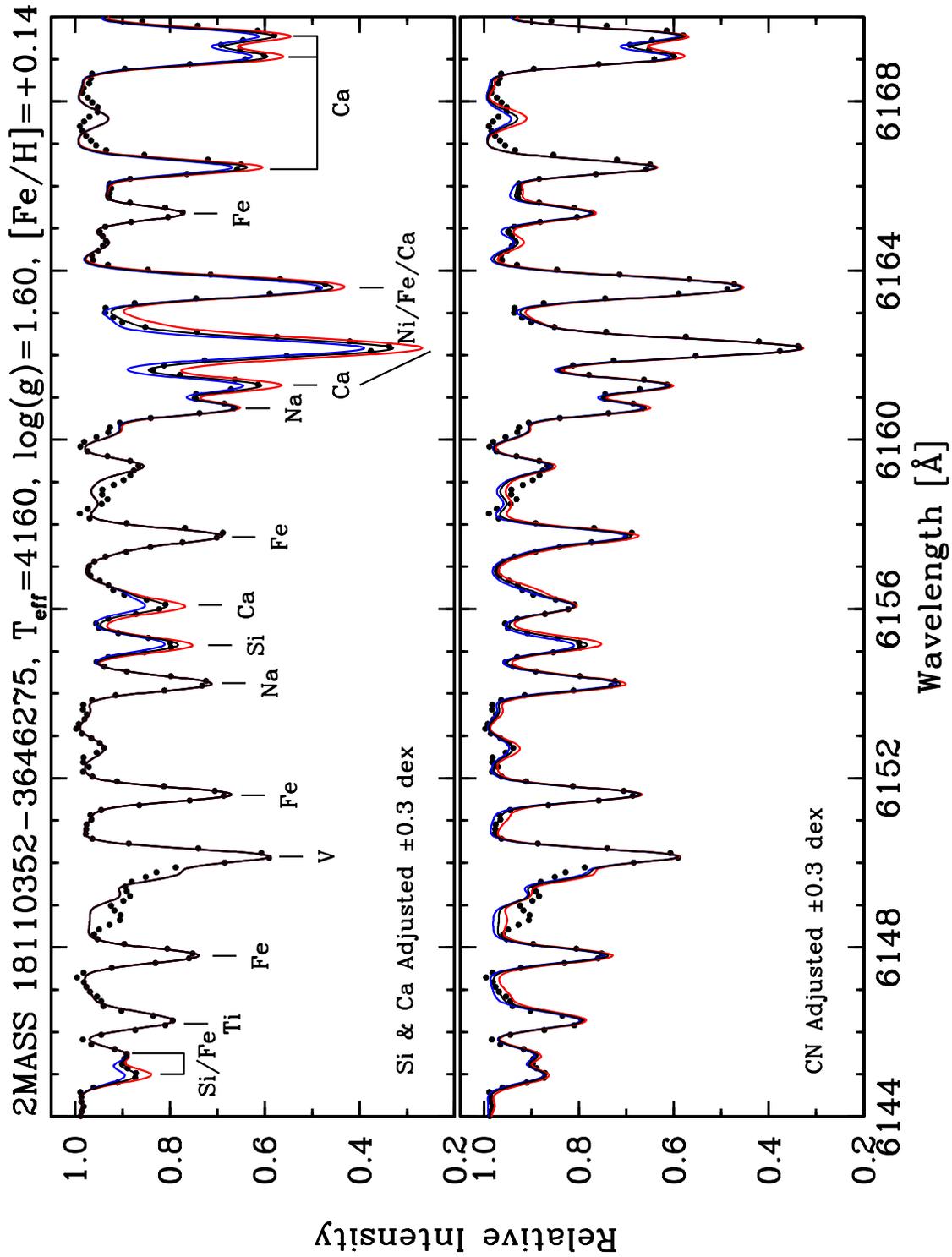}
\caption{Similar to Figure \ref{f4}, the synthetic spectrum fits shown here
illustrate how the Si and Ca line profiles change when the Si and Ca 
abundances are altered $\pm$0.3 dex from the best--fit value (top panel) and
when the CN abundance is altered by $\pm$0.3 dex from the best--fit value 
(bottom panel).  Note the particular sensitivity of the 6155 Si I line to 
changes in the CN abundance.}
\label{f6}
\end{figure}

The Si and Ca abundances are based on synthesis fits to lines in the 
6140--6170 \AA\ window.  In order to properly account for CN contamination,
Si and Ca were measured after obtaining CNO abundances.  For the few stars 
where [O I] could not be measured, we have adopted the general [O/Fe] versus 
[Fe/H] trend found in our data (see $\S$4.3).  Sample Si and Ca synthesis fits 
for a cool, metal--rich giant are shown in Figure \ref{f6}.  While most of the 
lines are not strongly affected by CN, the 6155 \AA\ Si I line is sensitive to 
the CN abundance in our most metal--rich giants.  As can be seen in 
Figure \ref{f6}, changing the CN abundance by $\pm$0.3 dex can result in a 
Si abundance uncertainty $\ga$0.2 dex.  A comparison between the 6155 \AA\ 
Si abundance and the 6145 \AA\ Si abundance, which is only weakly affected by 
CN, suggests that at least on average there is reasonable agreement with 
$\langle$[Si/Fe]$_{\rm 6155}$--[Si/Fe]$_{\rm 6145}$$\rangle$=--0.03 
($\sigma$=0.17).  The four Ca lines used here exhibit a similar degree of 
agreement with an average line--to--line dispersion of 0.08 dex 
($\sigma$=0.05).

\subsection{Abundance Uncertainty Estimates}

Tables 5a--5d summarize the abundance uncertainty estimates for 
log $\epsilon$(Fe), log $\epsilon$(O), log $\epsilon$(Si), and 
log $\epsilon$(Ca) due to changes in T$_{\rm eff}$$+$100 K, log(g)$+$0.30, 
[M/H]$+$0.30, and vt$+$0.30 km s$^{\rm -1}$.  The $\sigma$/$\surd$(N) values 
for each element are also given as an estimate of the measurement uncertainty. 
For elements in Tables 5b--5d where only one line was available for measurement 
(e.g., [O I]), a value of $\sigma$/$\surd$(N)=0.05 has been assigned.  This
represents the average value for cases where multiple lines could be measured,
and is a reasonable uncertainty estimate for visually fitting synthetic 
spectra.  The total error column in Tables 5a--5d reflects the T$_{\rm eff}$, 
log(g), [M/H], vt, and measurement uncertainties added in quadrature, and are 
reflected in the error bars of all subsequent figures.

The abundance sensitivities to each model atmosphere parameter were calculated
by varying each parameter independently.  The procedure involved first using 
the best--fit model atmosphere parameters and abundances given in Tables 2--3 
to calculate theoretical EWs for each line.  These theoretical EWs were held 
fixed and then a new model atmosphere was created with the T$_{\rm eff}$, 
log(g), [M/H], or vt value varied by the amount given in Tables 5a--5d.  
However, it should be noted that the sensitivity of each line to changes in
model atmosphere parameters are not necessarily the same.  Therefore, the 
values given in Tables 5a--5d represent the average difference between the 
newly derived abundances from each line and the original abundance given in
Table 2.

The uncertainty ranges of $\Delta$T$_{\rm eff}$=$+$100 K, 
$\Delta$log(g)=$+$0.30, $\Delta$[M/H]=$+$0.30, and 
$\Delta$vt=$+$0.30 km s$^{\rm -1}$ are likely to be conservative upper limits
for each parameter.  The T$_{\rm eff}$ variations, based on the uncertainty
in E(B--V), are typically small ($<$50 K) for almost all stars in our sample.  
In particular, the average differences between the T$_{\rm eff}$ derived 
assuming the recommended E(B--V) and the maximum/minimum E(B--V) values are:
$+$38 K ($\sigma$=27 K)/--24 K ($\sigma$=15 K) for the (--5.5,--7) field, 
$+$14 K ($\sigma$=7 K)/--10 K ($\sigma$=6 K) for the (--4,--9) field, and
$+$21 K ($\sigma$=12 K)/--22 K ($\sigma$=10 K) for the ($+$8.5,$+$9) field.
These values are significantly smaller than the 125 K 1$\sigma$ value of the 
Alonso et al. (1999) J--K$_{\rm S}$ color--temperature calibration.  When we 
compare the T$_{\rm eff}$ values derived using the Alonso et al. (1999) 
calibration to those derived using the Gonz{\'a}lez Hern{\'a}ndez \& Bonifacio 
(2009) calibration, we find our adopted Alonso et al. (1999) temperatures to be
cooler by 63 K ($\sigma$=8 K) in the (--5.5,--7) field, 68 K ($\sigma$=7 K) in 
the (--4,--9) field, and 46 K ($\sigma$=13 K) in the ($+$8.5,$+$9) field.  
Therefore, we believe that an estimated uncertainty of 100 K is a reasonable 
assumption for our data.

Since we did not calculate log(g) directly from photometry or ionization 
equilibrium, it is difficult to assess the true log(g) uncertainty.  However,
examination of Figure \ref{f2} suggests that the difference between our 
derived log(g) values and those in the literature is typically within 
$\pm$0.30.  For comparison, the Bescancon model\footnote{The Bescancon model
form can be accessed at: http://model.obs-besancon.fr/.} (Robin et al. 2003)
predicts that most bulge stars along our observed lines--of--sight should 
vary in distance by approximately $\pm$2 kpc, though the full range may extend
several kpc on either side of the Galactic center.  This would lead to 
a maximum photometric log(g) uncertainty of $\sim$0.20--0.25.  Note that 
oxygen, especially when normalized to Fe I, is the element most affected by 
log(g) uncertainties.  Similarly, since surface gravity is proportional to
log(M/M$_{\sun}$) for stars with the same T$_{\rm eff}$ but different mass, it
is likely that our abundance ratios will not be significantly altered by the
possible inclusion of lower mass AGB stars in our sample.

As stated previously, the average line--to--line dispersion for Fe I is 0.16 
dex, with a small dispersion of 0.03 dex.  Similarly, we found that the use of
$\alpha$--enhanced versus solar--scaled models affected abundance ratios at
$\la$0.1 dex level.  Therefore, it seems likely that the use of a model 
atmosphere [M/H] uncertainty of 0.3 dex should represent a conservative 
upper limit on the effects continuous opacity and electron number density have
on derived abundances.  From Tables 5a--5d, we find that [M/H] uncertainties
have the largest effect ($\sim$0.1 dex in magnitude) on O and Fe while Si and 
Ca abundances are less sensitive.  The magnitude of the [M/H] uncertainty 
effect also tends to increase with increasing metallicity.

Our adopted microturbulence uncertainty of 0.3 km s$^{\rm -1}$ may also be a 
conservative estimate.  For an individual star, the typical vt uncertainty 
based on removing any trends in Fe I abundance versus line strength is 
$\sim$0.1--0.15 km s$^{\rm -1}$.  When comparing stars of similar metallicity
(in 0.5 dex bins), the 1$\sigma$ variation in derived vt values for stars of
comparable temperature ranges from 0.17 km s$^{\rm -1}$ in the most metal--poor
stars to 0.22 km s$^{\rm -1}$ in the most metal--rich stars.  The effect of 
vt uncertainty on abundance is clear in Tables 5a--5d, which are sorted by
metallicity in each field.  Microturbulence becomes an increasingly important
factor with increasing metallicity and thus line strength.  Similarly, 
elements that were derived from stronger lines (e.g., Fe and Ca) are most 
sensitive to vt uncertainty.

\subsection{Radial Velocity Determinations}

Radial velocities for all stars were measured using the \emph{fxcor} task in
IRAF.  For templates, we calculated synthetic spectra over the full wavelength
range covered for each star.  The synthetic spectra were produced using the 
model atmosphere parameters given in Table 2, and were smoothed and binned
to match the dispersion of the object spectra.  The average measurement error 
returned by \emph{fxcor} was similar for all fields: 0.71 km s$^{\rm -1}$ 
($\sigma$=0.48 km s$^{\rm -1}$), 0.88 km s$^{\rm -1}$ ($\sigma$=0.53 km 
s$^{\rm -1}$), and 0.62 km s$^{\rm -1}$ ($\sigma$=0.38 km s$^{\rm -1}$) for the
(--5.5,--7), (--4,--9), and ($+$8.5,$+$9) fields, respectively.  The IRAF task
\emph{rvcorrect} was used the determine the heliocentric correction for all
stars.  The derived heliocentric radial velocities are provided in Table 2.

\section{RESULTS AND DISCUSSION}

\subsection{Metallicity Distribution Functions}

\begin{figure}
\epsscale{0.90}
\plotone{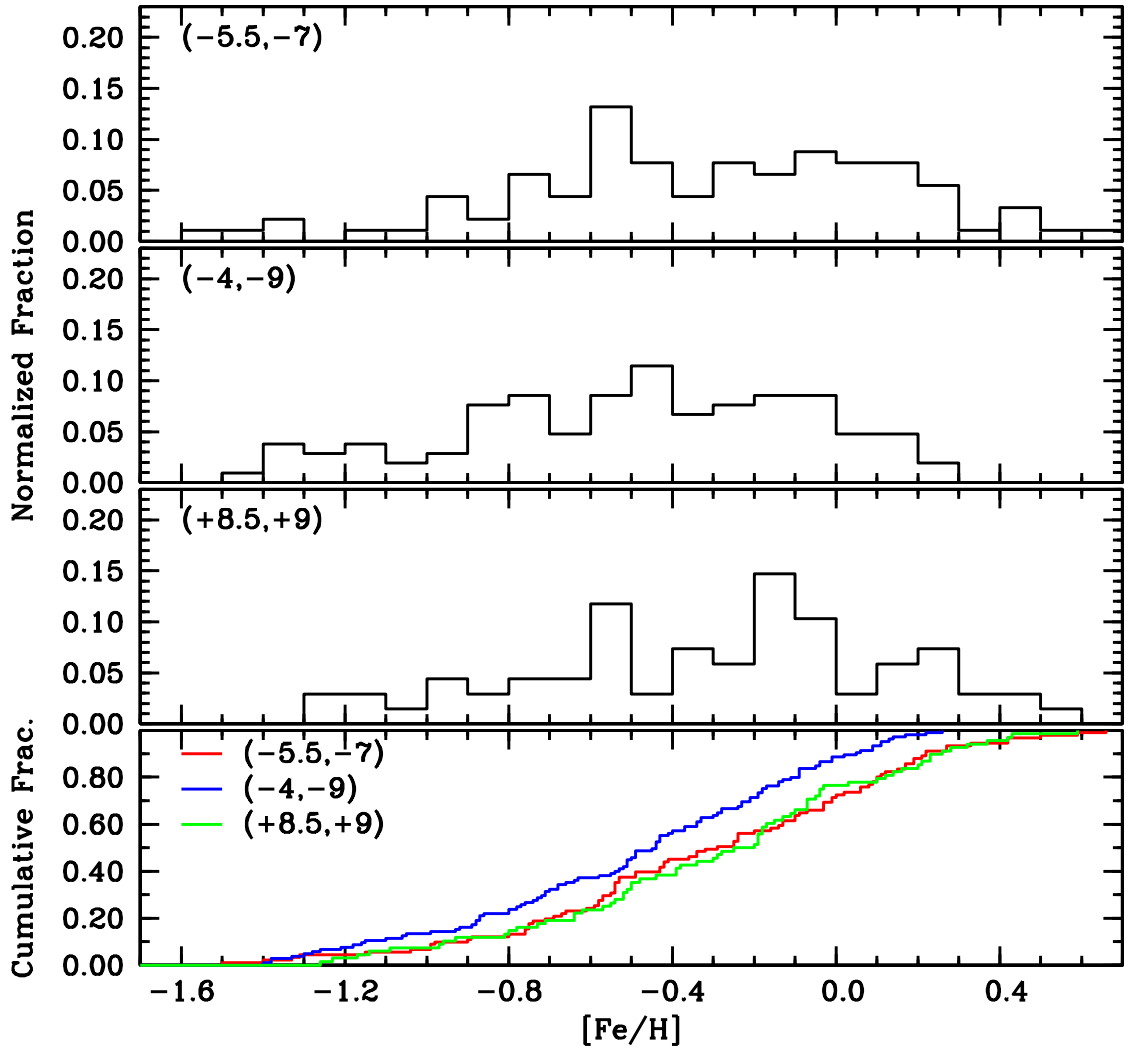}
\caption{The top three panels show the metallicity distribution functions 
for all three bulge fields binned in 0.1 dex increments.  The bottom panel
compares the cumulative distribution functions for all three fields.} 
\label{f7}
\end{figure}

In Figure \ref{f7} we show the derived metallicity distribution functions for
all three off--axis fields.  These metallicity distributions are in reasonable 
agreement with those derived for similar outer bulge 
($\mid$b$\mid$$>$4$\degr$) minor--axis fields (Zoccali et al. 2008; Johnson et 
al. 2011; Ness et al. 2013; Uttenthaler et al. 2012).  For the (--5.5,--7), 
(--4,--9), and ($+$8.5,$+$9) fields, we find a full range in [Fe/H] that spans 
--1.50 to $+$0.66 dex ($\sigma$=0.47 dex), --1.40 to $+$0.26 dex ($\sigma$=0.42
dex), and --1.26 to $+$0.59 dex ($\sigma$=0.44 dex), respectively.  As is the 
case with previous high resolution spectroscopic studies, we do not find any 
stars with [Fe/H]$\la$--1.5 that may be present in the bulge from the earliest 
epoch of star formation.  This is in agreement with the ARGOS sample by Ness et
al. (2013) that only found 0.71$\%$ (100/14147) of stars within 3.5 kpc of the 
Galactic center to have [Fe/H]$<$--1.5.

\begin{figure}
\epsscale{1.00}
\plotone{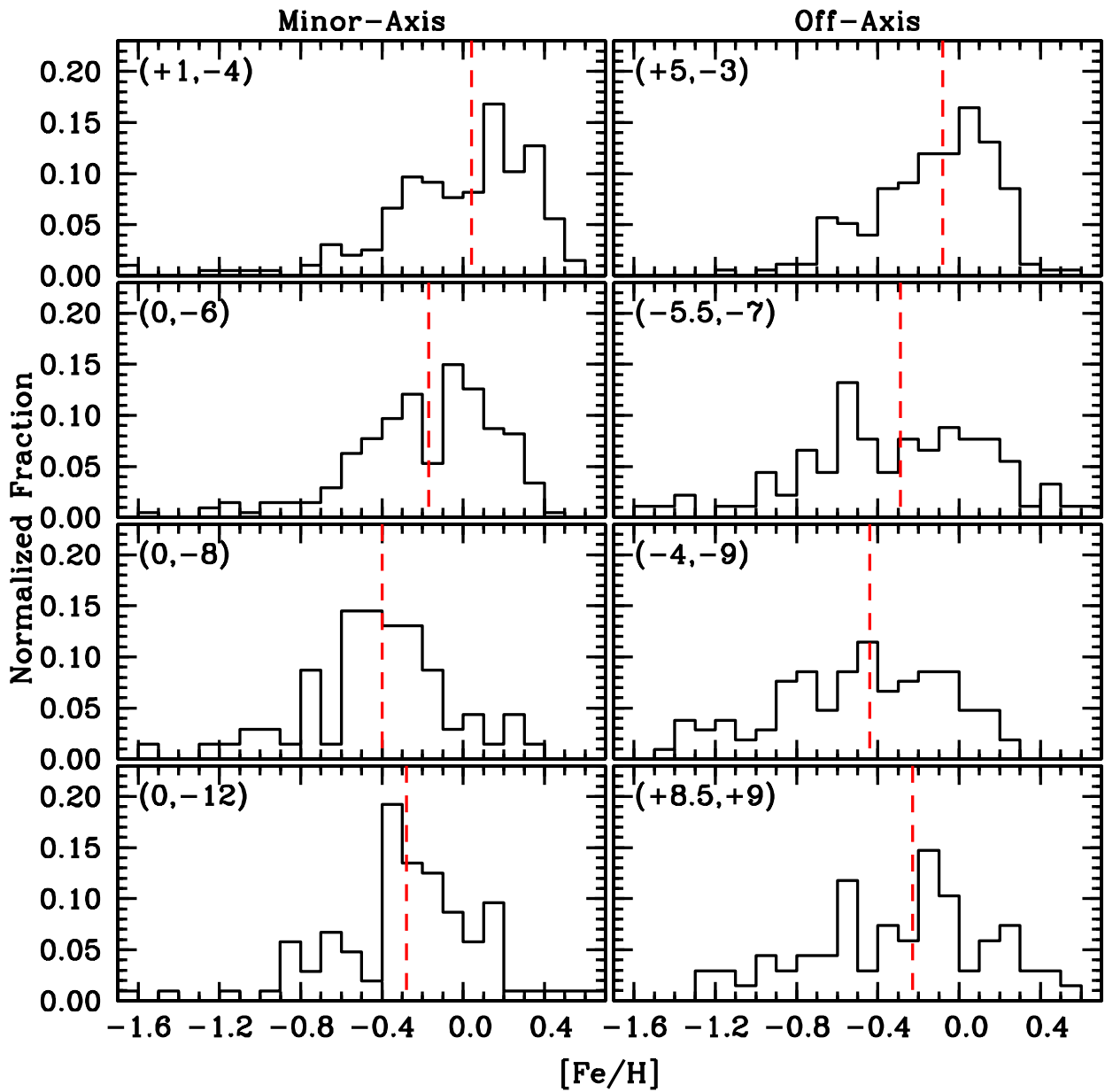}
\caption{Metallicity distribution functions are shown for bulge RGB stars in
four minor--axis (left panels) and four off--axis (right panels) fields.
The dashed red line in each panel indicates the median [Fe/H] value.  The
(l,b)=($+$1,--4), (0,--6), and (0,--12) fields are from Zoccali et al. (2008),
the (0,--8) field is from Johnson et al. (2011), the ($+$5,--3) field is from
Gonzalez et al. (2011), and the (--5.5,--7), (--4,--9), and ($+$8.5,$+$9)
fields are from the present work.  Note that the decreasing metallicity
gradient with increasing Galactic latitude appears to be present along both
the minor axis and off--axis.}
\label{f8}
\end{figure}

The observed vertical metallicity gradient present along the 
bulge minor--axis, at least for $\mid$b$\mid$$>$4$\degr$ (Zoccali et al. 2008), 
appears to also be present in off--axis fields.  This result is 
illustrated in Figure \ref{f8}, where we show the metallicity distribution 
functions for four minor--axis and four off--axis fields at similar Galactic 
latitudes.  In a similar fashion to the minor--axis trend, the median [Fe/H] 
ratios for the ($+$5,--3) field given in Gonzalez et al. (2011) and our 
(--5.5,--7) and (--4,--9) fields are [Fe/H]=--0.08, --0.29, and --0.44, 
respectively.  The decline in the median [Fe/H] ratio with increasing Galactic 
latitude appears to be driven by the same phenomenon in both the minor--axis 
and off--axis fields.  In particular, both sets of observations show a decrease 
in the relative fraction of metal--rich ([Fe/H]$\ga$0) stars, an increase
in the relative fraction of the metal--poor ([Fe/H]$\la$--1) stars, and no
significant change in the relative fraction of intermediate metallicity stars
as a function of increasing Galactic latitude.  This result is at least 
qualitatively in agreement with the scenario proposed by Babusiaux et al. 
(2010), in which the metallicity gradient is a reflection of the change in the 
bulge/bar population mixture at different Galactic latitudes (but see also 
Ness et al. 2013).

The data presented above indicate that the vertical metallicity gradient along 
l$\sim$$\pm$5$\degr$ is approximately 0.4 dex kpc$^{\rm -1}$, which is similar 
to the minor--axis gradients found by Zoccali et al. (2008; 0.6 dex 
kpc$^{\rm -1}$) and Ness et al. (2013; 0.45 dex kpc$^{\rm -1}$)\footnote{Note 
that Ness et al. (2013) also argue that the metallicity gradient of the 
individual bulge components they have identified may be significantly more 
shallow, on the order of 0.07 dex kpc$^{\rm -1}$, than the 0.4--0.6 dex 
kpc$^{\rm -1}$ found here and in previous studies.  Unfortunately, our sample 
sizes are insufficient to attempt a similar multi--component deconvolution of 
the distribution functions shown in Figure \ref{f7}}.  Given the uncertainty in
bulge membership and possible biases of incomplete samples, it seems unlikely 
that the difference in magnitude ($\sim$0.2 dex) between the derived 
metallicity gradients found here and in Zoccali et al. (2008) is significant. 
Additionally, there does not seem to be a strong radial [Fe/H] gradient in
the bulge. 

Interestingly, the northern bulge ($+$8.5,$+$9) field is slightly more 
metal--rich than the two southern fields, with a median [Fe/H]=--0.23.  
However, as can be seen in Figure \ref{f7}, where we plot the cumulative [Fe/H] 
distribution functions for all fields, it is clear that the the ($+$8.5,$+$9) 
and (--5.5,--7) fields share similar distributions.  This is supported by the 
results of two--sided Kolmogorov--Smirnov (KS) tests (Press et al. 1992), 
which indicate that there is insufficient evidence to reject the null 
hypothesis that the ($+$8.5,$+$9) and (--5.5,--7) are drawn from the same
parent population (p--value=0.881)\footnote{We adopt the common interpretation
that the null hypothesis can be rejected if p$<$0.05.}.  On the other hand,
the (--5.5,--7)/(--4,--9) and (--4,--9)/($+$8.5,$+$9) field combinations are
possibly drawn from different parent populations, with p--values of 0.031 and
0.045, respectively.  It is not clear if this signifies an asymmetry in the 
bulge metallicity distribution between the northern and southern regions or is
simply a result of the smaller sample size and slightly redder colors of the
observed stars in the ($+$8.5,$+$9) field (see Figure \ref{f1}).  We also 
consider the possibility that the higher metallicity of the 
($+$8.5,$+$9) field could be due to a predominance of the X--shaped bulge 
structure in that region.  Ness et al. (2012) and Uttenthaler et al. (2012)
suggest that the X--shaped bulge, traced by the double red clump, may be 
dominated by metal--rich ([Fe/H]$>$--0.5) stars.  As can be seen in Saito et 
al. (2011; their Figure 3), one of the red clump components in the X--shaped 
structure appears prominently near our ($+$8.5,$+$9) sight line.  A distant, 
lower density component is also noted near the (--5.5,--7) sight line, and the 
higher metallicity of both our (--5.5,--7) and ($+$8.5,$+$9) could partially
reflect the predominance of the X--shaped structure in these fields.

\subsubsection{Sample Selection Bias and Foreground Contamination}

As mentioned in $\S$2.1, the targets in all three bulge fields were selected 
from the blue side of the RGB to avoid cool stars with spectra that may be 
dominated by molecular bands (especially TiO).  The naive interpretation is 
that this will produce a strong bias in our derived metallicity distribution 
functions and significantly undersample the most metal-rich stars.  However,
the bias may not be as strong as expected because several factors combine to 
disperse the observed J--K$_{\rm S}$--[Fe/H] relation including: differential 
reddening, a complicated spatial geometry with a distance spread that could 
range from $\sim$4.5--11.5 kpc in our fields\footnote{This estimate is based 
on the Bescancon model.}, metallicity dependent [$\alpha$/Fe] variations, a 
possible age range of $\sim$2--13 Gyr for stars with [Fe/H]$\ga$--0.4 (e.g., 
see Figures 15--16 of Bensby et al. 2013; but see also Clarkson et al. 2008), 
and the mixing of RGB and AGB stars.

\begin{figure}
\epsscale{1.00}
\plotone{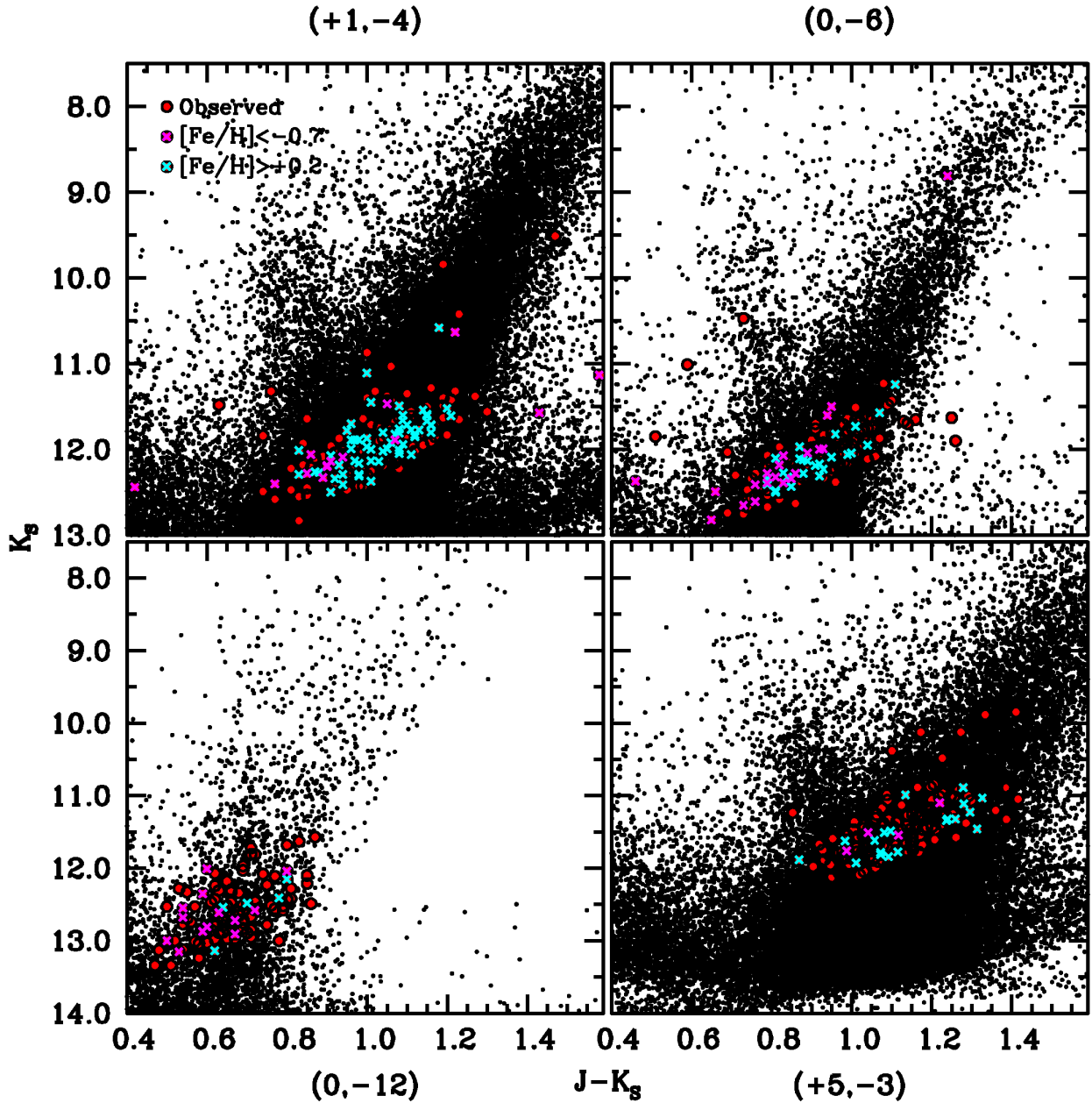}
\caption{Color--magnitude diagrams using 2MASS data (small black points) are 
shown for the bulge fields presented in Zoccali et al. (2008) and 
Gonzalez et al. (2011).  The filled red circles are the observed stars in 
each field, those with magenta crosses are the most metal--poor stars 
([Fe/H]$<$--0.7), and those with cyan crosses are the most metal--rich stars 
([Fe/H]$>$$+$0.2).  Note that both metal--poor and metal--rich stars are found 
over a broad range in J--K$_{\rm S}$ color.}
\label{f9}
\end{figure}

In general, it is not a trivial matter to estimate the observational 
bias of the color--metallicity relation in the bulge, especially in the 
J--K$_{\rm S}$ and K$_{\rm S}$ dimensions, because of large uncertainties in 
several of the factors listed above.  This is evident both in the metallicity 
distribution functions produced here (Figure \ref{f7}), which are not 
dominated by metal--poor stars as one might expect, and in the 
J--K$_{\rm S}$--[Fe/H] relations of previous studies that span a broader color 
range.  In Figure \ref{f9} we show K$_{\rm S}$ versus 
J--K$_{\rm S}$ color--magnitude diagrams for the Zoccali et al. (2008) and 
Gonzalez et al. (2011) data sets and overplot the observed stars, with the most
metal--poor ([Fe/H]$<$--0.7) and metal--rich ([Fe/H]$>$$+$0.2) subsets 
identified.  While the differential reddening and distance distributions vary 
from field--to--field, a common feature among all fields is the significant 
mixing of metal--poor and metal--rich stars on the giant branch.  The 
metal--rich population in particular appears to span the entire color range, 
although an unknown fraction of these stars may be disk and/or halo 
contaminators (see below).

\begin{figure}
\epsscale{1.00}
\plotone{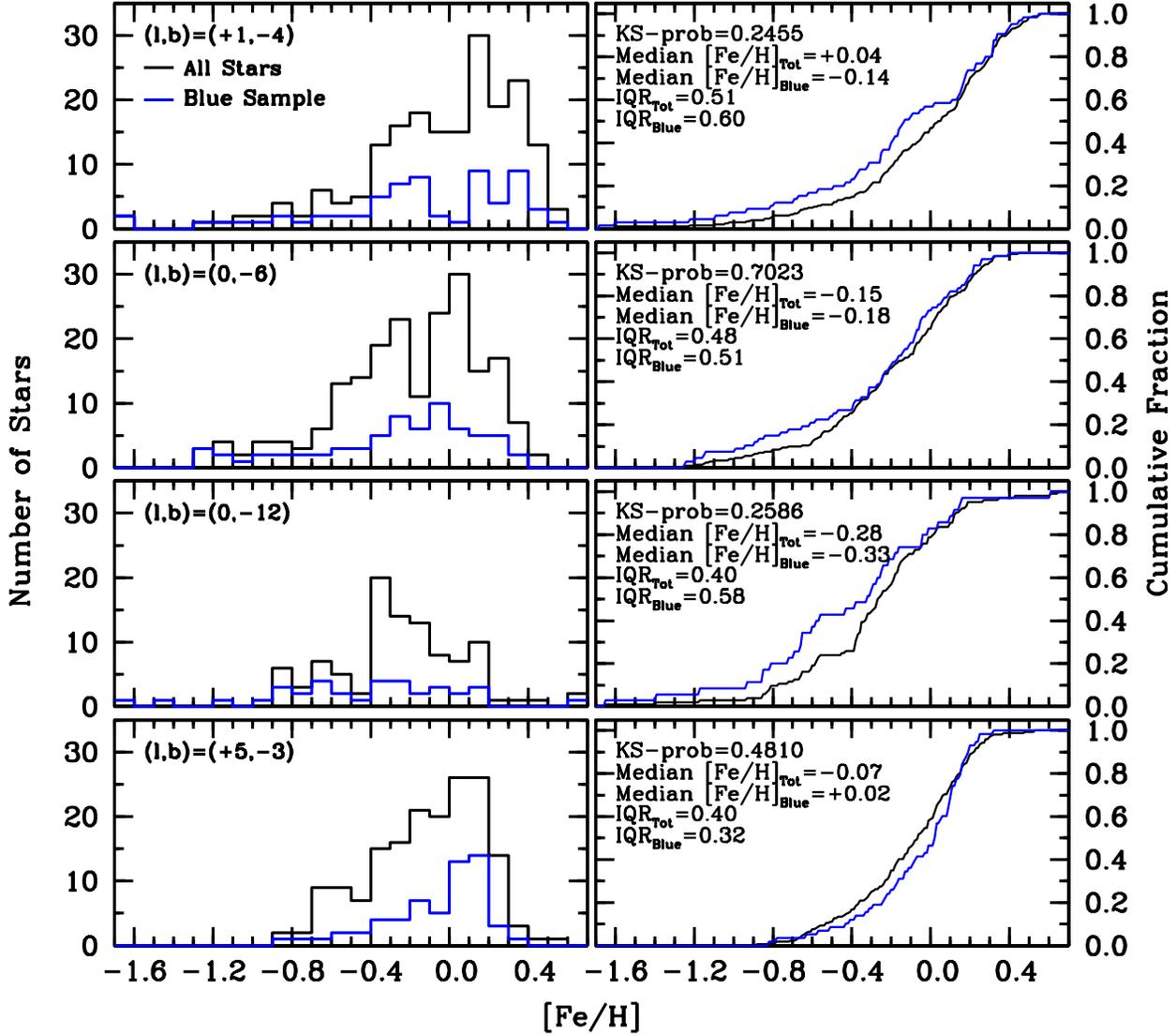}
\caption{Metallicity distribution histograms and cumulative distribution 
functions are shown for the full samples (solid black lines) presented in 
Zoccali et al. (2008) and Gonzalez et al. (2011).  Similar data are plotted 
using just the bluest third of each sample (solid blue lines), which roughly
matches our sample selection.  In the right panels we include the results of 
two--sided KS tests (comparing the blue and total samples), median [Fe/H] 
values, and interquartile ranges (IQR) for each population.}
\label{f10}
\end{figure}

What is the effect on the derived metallicity distribution functions by 
preferentially sampling the blue side of the bulge giant branch?  We can derive
some estimate with the Zoccali et al. (2008) and Gonzalez et al. (2011) fields
by comparing the metallicity distribution functions of the bluest third, which
roughly matches our sample selection, to the full samples.  The 
results are shown in Figure \ref{f10} where we plot histograms, cumulative 
distribution functions, provide results for two--sided KS tests, compare the 
median [Fe/H] values, and compare the inter--quartile range (IQR) for each
population.  While it is clear from the cumulative distribution functions that
(in most cases) the blue samples are moderately more metal--poor than the 
full samples, the effect is significantly smaller than anticipated.  In fact,
the two--sided KS tests for all fields indicate that we cannot strongly reject 
the null hypothesis that the blue and full samples were drawn from the same 
parent population.  Furthermore, the general shapes of the full metallicity 
distribution functions are preserved in the blue samples, and the median [Fe/H]
and IQR values do not vary significantly between the two subsets, especially
in the b$>$--4$\degr$ fields that are most similar to our fields.  In other 
words, the qualitative and quantitative results found especially by
Zoccali et al. (2008), and particularly with regard to the [Fe/H] gradient, 
would not have been significantly altered by the selection of stars on the blue
side of the giant branch\footnote{We tested for the effects of differential 
reddening and photometric errors in the 2MASS J and K$_{\rm S}$ data but did 
not find significant changes to the distribution functions or median [Fe/H] 
and IQR values.}.  

Finally, we can compare our (--5.5,--7) and (--4,--9) fields with the data 
presented in Ness et al. (2013; their Figure 8).  For the similar latitude 
minor--axis fields of (0,--7.5; 690 stars) and (0,--10; 650 stars) in Ness et 
al. (2013), they find median [Fe/H] values of --0.34 and --0.46, respectively.
This is in good agreement with the median [Fe/H] values of our (--5.5,--7) and 
(--4,--9) fields, which have [Fe/H]=--0.29 and -0.44, respectively.  The 
$<$0.05 dex difference between our derived median [Fe/H] values and Ness et al.
(2013) is in agreement with the estimated bias introduced by the selection of 
stars on the blue side of the giant branch (see the middle panels of Figure 
\ref{f10}).  We conclude that our selection of bluer stars likely does not
introduce a strong metallicity bias, and that the derived metallicity
distribution functions are representative of the underlying population in each 
field.  However, the metal--rich stars in our sample will be more concentrated
on the near side of the bulge/bar system than a selection of stars with a 
broader color distribution.

\begin{figure}
\epsscale{1.00}
\plotone{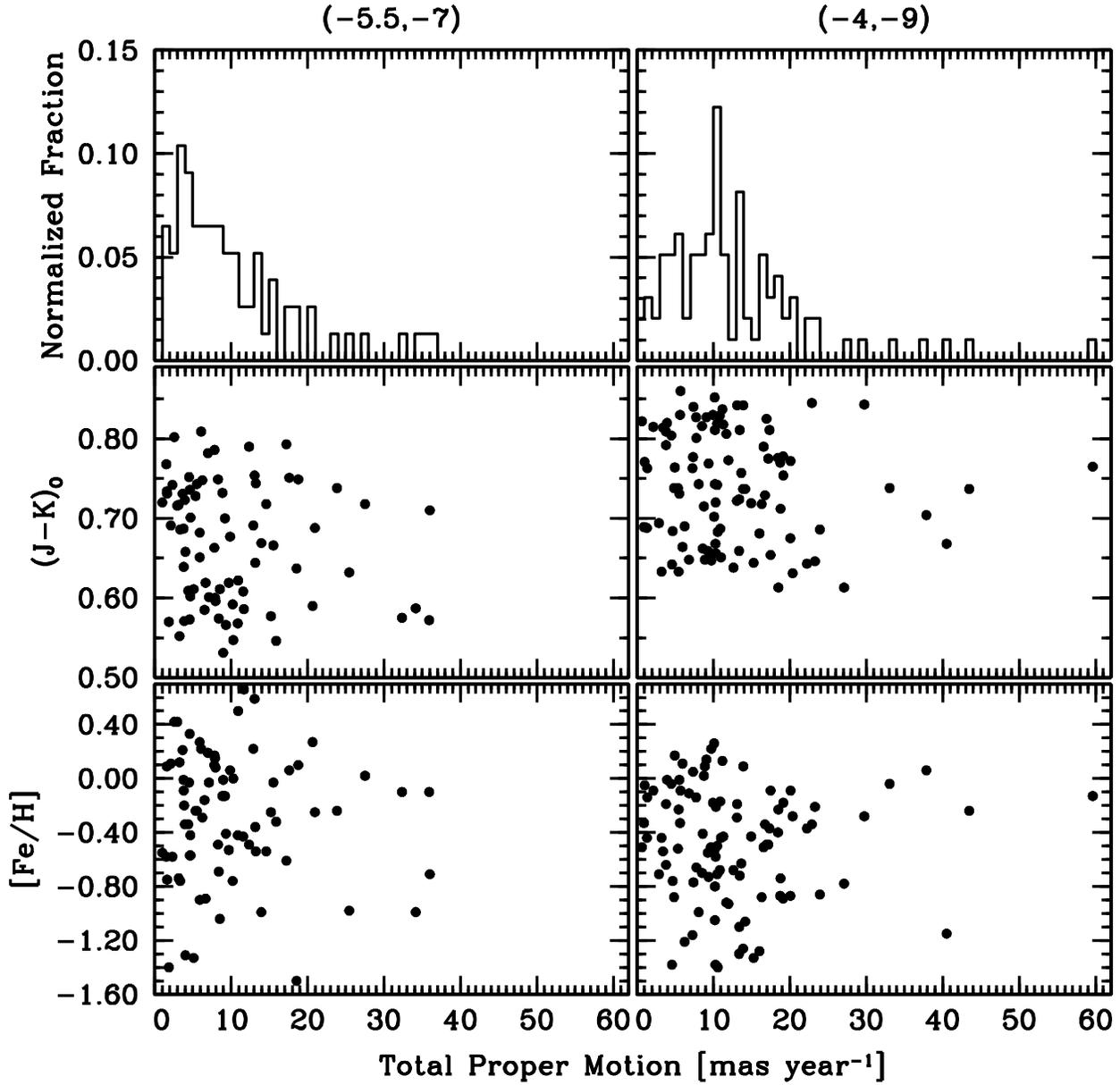}
\caption{The left panels show (top) a histogram of the total proper motion 
for stars in the (--5.5,--7) field, (middle) a plot of (J--K)$_{\rm o}$ versus
total proper motion, and (bottom) a plot of [Fe/H] versus total proper motion. 
The right panels include the same information for the (--4,--9) field.  The 
proper motion data are from the SPM4 catalog.}
\label{f11}
\end{figure}

In addition to the properties mentioned above that can blur the 
color--metallicity relation, contamination by disk and halo stars along the 
line--of--sight can also affect the derived metallicity distribution functions.
While most studies agree that the halo contamination is small ($\la$2$\%$; 
e.g., Zoccali et al. 2008; Hill et al. 2011; Uttenthaler et al. 2012), a true
estimate of the (primarily) foreground thin and thick disk contamination
is difficult to ascertain.  In an attempt to gauge the impact of foreground
contamination, the coordinates of stars in our two southern bulge fields were
cross--referenced with the Southern Proper Motion IV catalog (SPM4; Girard
et al. 2011)\footnote{The stars in our northern bulge field have a declination
of $\sim$--16$\degr$, which is outside the SPM4 range of --20$\degr$ to
--90$\degr$.}.  We were able to match 77/91 stars in the (--5.5,--7) field and
98/105 stars in the (--4,--9) field.  Histograms of the proper motion results
are shown in the top two panels of Figure \ref{f11}.  As can be seen in
Figure \ref{f11}, both fields contain a dominant population with a total proper
motion $\la$20 mas year$^{\rm -1}$ and a tail that extends out to $\sim$35 mas
year$^{\rm -1}$ for the (--5.5,--7) field and $\sim$60 mas year$^{\rm -1}$ for
the (--4,--9) field.  This is in good agreement with that found by Uttenthaler
et al. (2012; their Figure 3) at (0,--10).

As a rough estimate of contamination, we can assign targets lying in the tails
of the proper motion distributions as possible foreground stars.  Using a
discriminator of 22 mas year$^{\rm -1}$ for the (--5.5,--7) field and 21 mas
year$^{\rm -1}$ for the (--4,--9) field suggests foreground contamination
rates of 9$\%$ (7/77) and 11$\%$ (11/98), respectively.  However, given that
many of the ``outliers" fall within a few mas year$^{\rm -1}$ of our cut--off
and the proper motion errors are generally $\sim$3--8 mas year$^{\rm -1}$, it
is possible that as many as 50$\%$ of the possible outliers are actually
bulge members.  Additionally, since Uttenthaler et al. (2012) find that
the double red clump and dual peaked metallicity distribution function features
are present in their sample of stars with proper motions $>$20 mas
year$^{\rm -1}$, we follow their assessment and have not excluded any stars in
our sample.  We do note however that the middle and bottom panels of
Figure \ref{f11}, which plot (J--K)$_{\rm o}$ and [Fe/H] versus total proper
motion, give some indications that the more extreme proper motion outliers may
clump together at colors and metallicities consistent with the foreground thin
disk and metal--poor thick disk (or metal--rich halo).

\subsection{Radial Velocities}

\begin{figure}
\epsscale{1.00}
\plotone{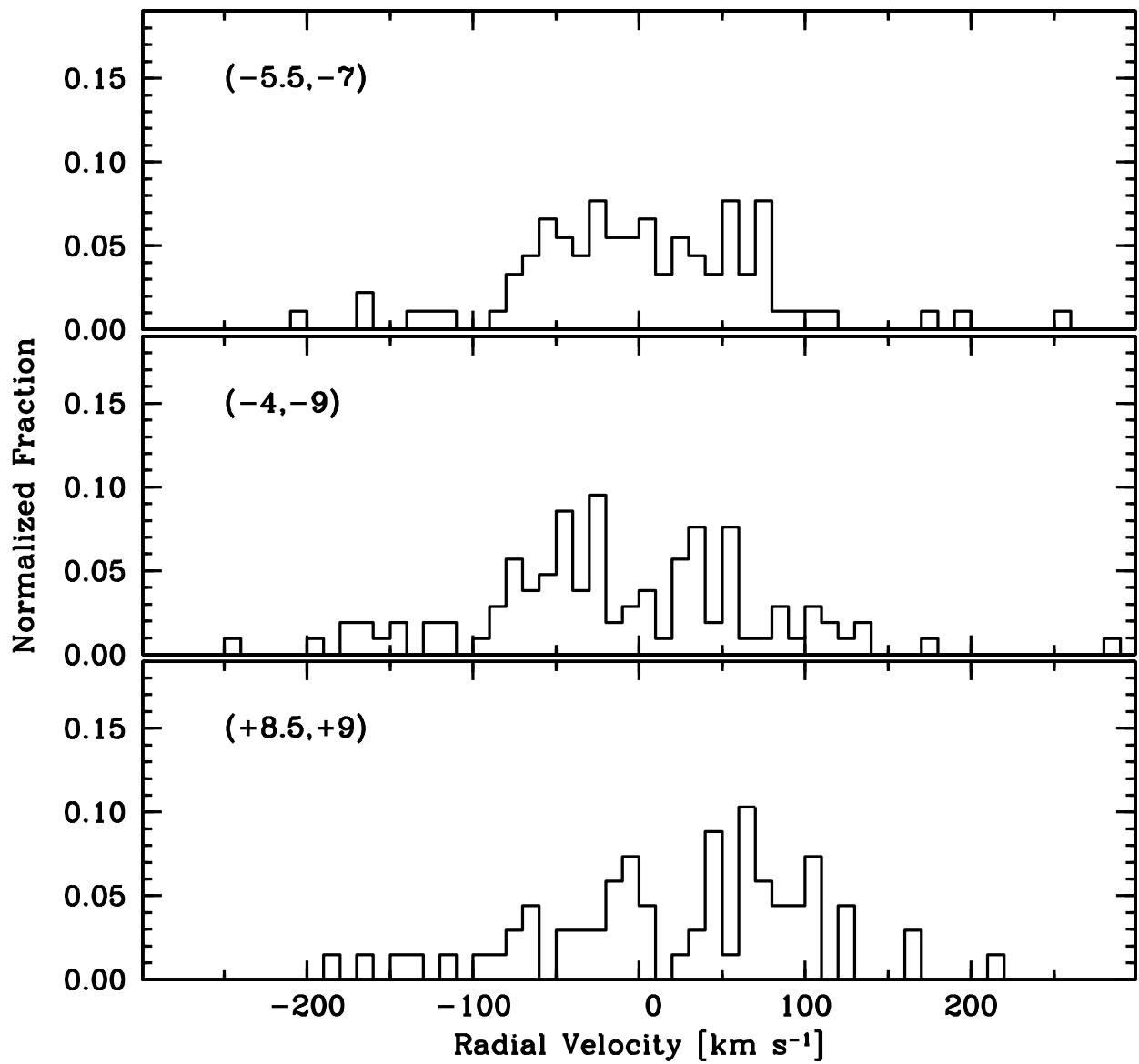}
\caption{The three panels show the heliocentric radial velocity distributions 
for each field in 10 km s$^{\rm -1}$ bins.}
\label{f12}
\end{figure}

\begin{figure}
\epsscale{1.00}
\plotone{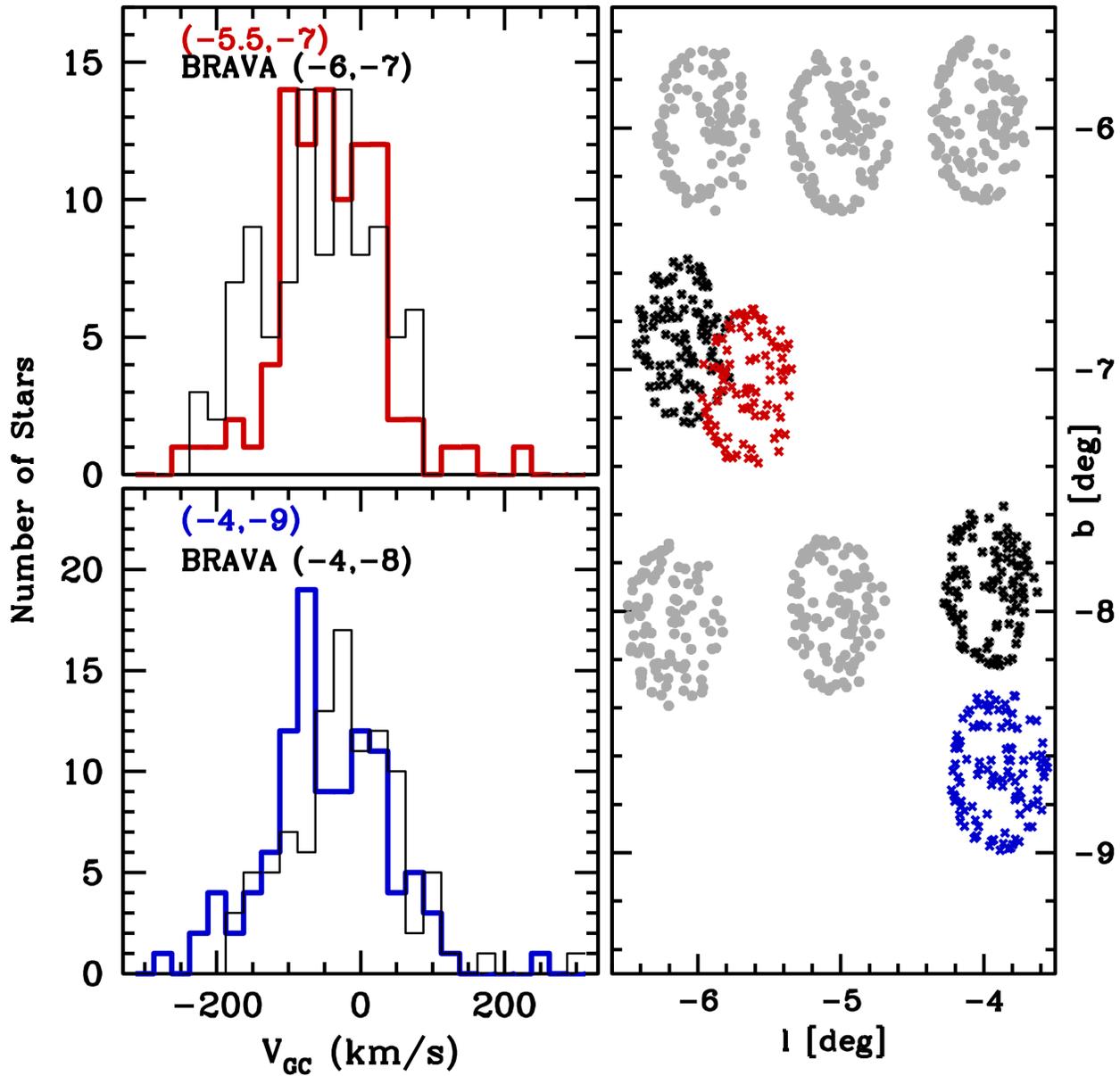}
\caption{The top left and bottom left panels compare the (--5.5,--7; red lines
and symbols) and (--4,--9; blue lines and symbols) to nearby and/or 
overlapping fields observed for the BRAVA project (Howard et al. 2008; Kunder 
et al. 2012).  The data are binned in 25 km s$^{\rm -1}$ increments.  The right
panel shows the Galactic longitude and latitude for individual stars in the 
new fields observed here, the comparison BRAVA fields used in the histograms 
(black crosses), and other nearby BRAVA fields (grey filled circles).}
\label{f13}
\end{figure}

In Figure \ref{f12} we plot histograms of the heliocentric radial velocity
distributions for all fields analyzed here.  The mean, median, and standard 
deviations are --26.6, --29.9, and 73.7 km s$^{\rm -1}$ for the (--5.5,--7) 
field, --40.7, --49.3, and 85.1 km s$^{\rm -1}$ for the (--4,--9) field, and 
--3.8, $+$15.2, and 81.6 km s$^{\rm -1}$ for the ($+$8.5,$+$9) field.  
Similarly, the mean/median galactocentric velocities (V$_{\rm GC}$) for these 
fields are: --40.8/--44.8 km s$^{\rm -1}$, --48.0/--57.8 km s$^{\rm -1}$, and 
$+$40.5/$+$60.2 km s$^{\rm -1}$.  For the two southern bulge fields, the 
velocity distributions and dispersions are in good agreement with the 
BRAVA results (e.g., see Kunder et al. 2012; their Figure 11).  This is 
confirmed in Figure \ref{f13} where we plot the velocity distributions from 
this work and overlapping or nearby BRAVA fields.  A two--sided KS test finds 
that the data are insufficient to strongly reject the null hypothesis that the 
(--5.5,--7)/(--6,--7) fields (p--value=0.063) and (--4,--9)/(--4,--8) fields 
(p--value=0.094) are drawn from the same parent populations.  Additionally,
the median galactocentric radial velocity and dispersion for our two southern
bulge fields are in reasonable agreement with the Shen et al. (2010) bulge 
model, which find V$_{\rm GC}$=--52.2 ($\sigma$=78.9) km s$^{\rm -1}$ and 
V$_{\rm GC}$=--54.1 ($\sigma$=78.2) km s$^{\rm -1}$ at (l,b)=(--5.5,--7) and 
(--4,--9), respectively (Z. Li \& J. Shen 2012; priv. comm.).

While the BRAVA sample did not extend to our single northern bulge field, the 
mean galactocentric velocity and dispersion we derived for the ($+$8.5,$+$9) 
field (V$_{\rm GC}$=$+$40.5 km s$^{\rm -1}$, $\sigma$=81.6 km s$^{\rm -1}$) is 
in moderately good agreement with that found by Minitti (1996) for the 
nearby ($+$8,$+$7) field (V$_{\rm GC}$=$+$54.4 km s$^{\rm -1}$, $\sigma$=84.4
km s$^{\rm -1}$).  Similarly, our derived median galactocentric velocity for
this field (V$_{\rm GC}$=$+$60.2 km s$^{\rm -1}$) is in reasonably good 
agreement with the Shen et al. (2010) median galactocentric velocity model 
value of V$_{\rm GC}$=$+$71.6 km s$^{\rm -1}$ ($\sigma$=65.7 km s$^{\rm -1}$; 
Z. Li \& J. Shen 2012; priv. comm.).  As was mentioned in $\S$4.1, it is 
possible that the ($+$8.5,$+$9) sight line may be partially contaminated by 
one of the structures associated with the X--shaped bulge (e.g., see Saito et 
al. 2011; their Figure 3).  This could result in additional substructure being 
present in our empirical velocity distribution function.  However, visual 
inspection of Figure \ref{f12} does not indicate strong evidence of any 
additional features.  This is supported by the results of Shapiro--Wilk and 
Anderson--Darling normality tests, which return p--values of 0.288 and 0.114, 
and indicate insufficient evidence to reject the null hypothesis that the 
($+$8.5,$+$9) velocity distribution is normal.

\begin{figure}
\epsscale{0.80}
\plotone{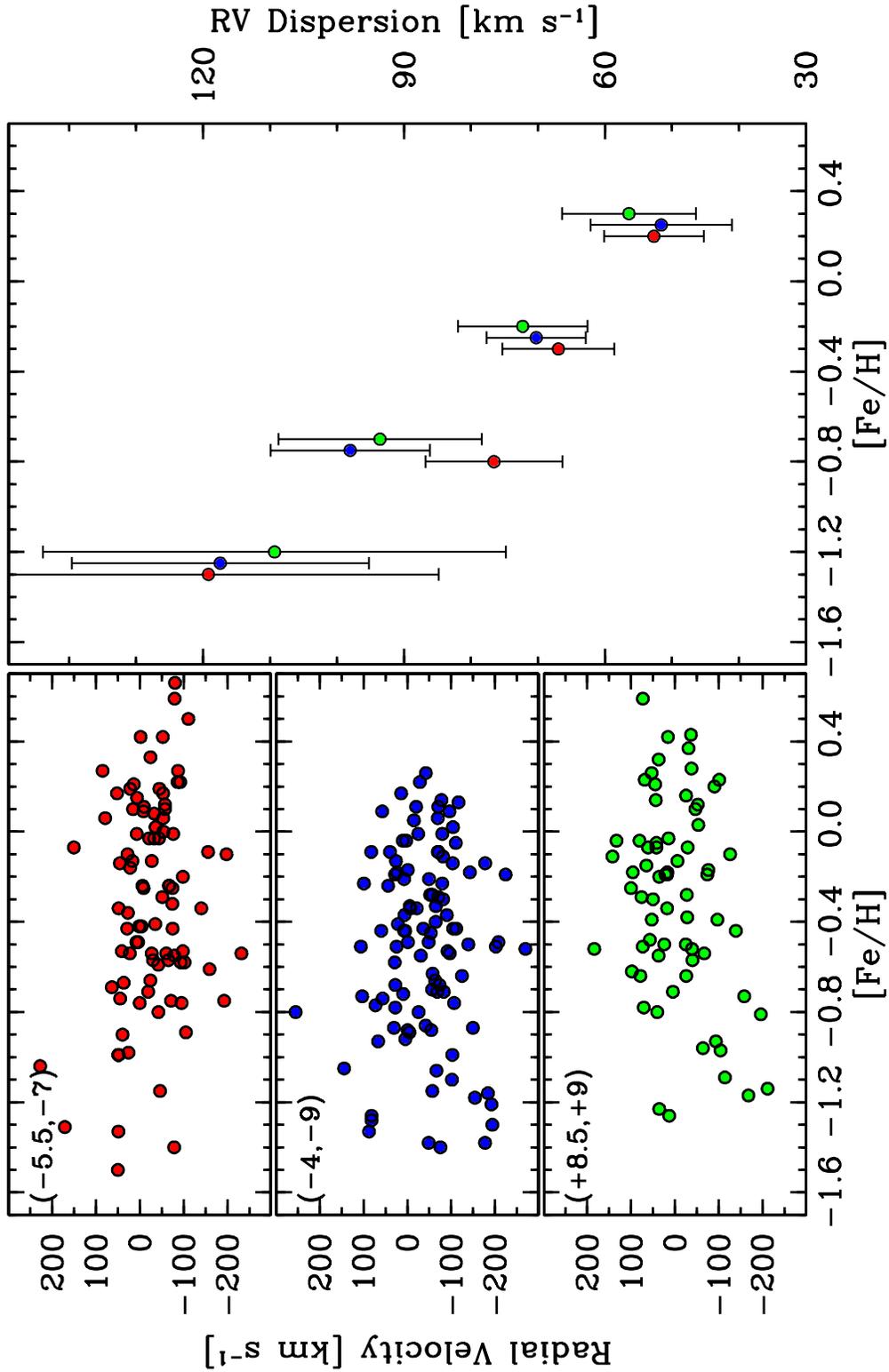}
\caption{The left panels show the heliocentric radial velocity distribution as 
a function of [Fe/H] for the (--5.5,--7) (filled red circles), (--4,--9) 
(filled blue circles), and ($+$8.5,$+$9) (filled green circles) fields.  In
the right panel, the velocity dispersion is shown for all three fields with
the data merged into 0.5 dex [Fe/H] bins.  The larger error bars on the most 
metal--poor bin are due to the relatively small sample size.}
\label{f14}
\end{figure}

In Figure \ref{f14} we show the heliocentric radial velocity and dispersion
distributions as a function of [Fe/H] for all fields.  As can be clearly seen
in Figure \ref{f14}, all three fields exhibit nearly identical trends of a 
decreasing velocity dispersion with increasing [Fe/H].  This observed trend is 
consistent with both the red clump and giant data analyzed in other outer
bulge ($\mid$b$\mid$$\ga$6$\degr$) fields (e.g., Minitti 1996; Babusiaux et al. 
2010; de Propris et al. 2011; Johnson et al. 2011; Uttenthaler et al. 2012), 
and contrasts with the minor--axis results found by Babusiaux et al. (2010; 
their Figure 6) at b=--4$\degr$ (opposite trend) and b=--6$\degr$ (no trend).
While de Propris et al. (2011) also find a correlation between the mean
radial velocity and [Fe/H] for red clump stars, Babusiaux et al. (2010) and 
Uttenthaler et al. (2012) do not find significant evidence of mean velocity 
variations as a function of metallicity.  Visual inspection of Figure \ref{f14}
indicates that we also do not find significant variations in mean/median 
velocity as a function of [Fe/H] with two exceptions: (1) the metal--poor
([Fe/H]$<$--1) stars in all fields and (2) the intermediate metallicity 
(--0.5$\la$[Fe/H]$\la$0) stars in the ($+$8.5,$+$9) field.  

When the data are partitioned into 0.5 dex [Fe/H] bins, the median radial 
velocities within each bin of the (--5.5,--7) and (--4,--9) fields vary by 
$<$20 km s$^{\rm -1}$ (ignoring the [Fe/H]$<$--1 stars).  The same is also true
in the ($+$8.5,$+$9) field where the stars with --1$<$[Fe/H]$<$--0.5 and 
[Fe/H]$>$0 have median radial velocities of --24.6 and --28.4 km s$^{\rm -1}$, 
respectively.  However, the stars in this field with --0.5$<$[Fe/H]$<$0 exhibit
a noticeable velocity increase to $+$28.8 km s$^{\rm -1}$.  Although there is 
evidence that the X--shaped bulge structure traced by the double red clump may 
preferentially contain stars with [Fe/H]$>$--0.5 (Ness et al. 2012; Uttenthaler
et al. 2012), it is not clear if the increase in median velocity near 
[Fe/H]$\sim$--0.5 for our RGB sample is due to contamination from the nearby 
X--shape structure or is simply a statistical fluctuation. 

For the metal--poor ([Fe/H]$<$--1) stars in the (--5.5,--7), (--4,--9), and 
($+$8.5,$+$9) fields the median heliocentric velocities are: $+$49.6 
($\sigma$=119.2) km s$^{\rm -1}$, --70.8 ($\sigma$=117.4) km s$^{\rm -1}$, and 
--114 ($\sigma$=109.3) km s$^{\rm -1}$, respectively.  This contrasts with the
more homogeneous set of stars with [Fe/H]$>$--1, which have median velocities
and dispersions of: --32.5 ($\sigma$=66.1) km s$^{\rm -1}$, --42.1 
($\sigma$=79.4) km s$^{\rm -1}$, and $+$17.4 ($\sigma$=76.1) km s$^{\rm -1}$.
However, the relatively small number of stars in the most metal--poor bin of 
each field makes it difficult to assess the significance of the larger median 
velocities and dispersion of these stars.  Therefore, we instead attempt to 
determine how many of the stars with [Fe/H]$<$--1 might be outliers by using
the interquartile range (IQR) of the more metal--rich ([Fe/H]$>$--1) stars
in each field.  We define moderate outliers as stars with velocities of 
1.5--3.0$\times$IQR above/below the median and extreme outliers as stars
having velocities $>$3.0$\times$IQR above/below the median.  While we did not
find extreme outliers in any of the fields, $\sim$20--40$\%$ of the metal--poor
stars are moderate outliers and may belong to the halo.  The small percentage
of stars with both low [Fe/H] and high velocities is consistent with the
halo contamination estimate given in $\S$4.1.1.

\subsection{$\alpha$--Element Abundances}

While iron may be produced by both short lived Type II supernovae (SNe) and 
longer lived Type Ia SNe, the $\alpha$ elements are primarily produced only by 
massive stars.  As a consequence, the [$\alpha$/Fe] ratio can be a useful 
diagnostic for determining the chemical enrichment history of a stellar 
population (e.g. Tinsley 1979; Matteucci \& Brocato 1990).  In particular, the 
magnitude of the [$\alpha$/Fe] ratios and the metallicity at which the
[$\alpha$/Fe] ratio begins to decline (presumably due to the onset of Type Ia 
contributions) yield information about important quantities such as a 
population's initial mass function (IMF) and star formation rate (e.g., see 
review by McWilliam 1997 and references therein).  Additionally, the 
star--to--star dispersion of the measured [$\alpha$/Fe] ratios in a given
metallicity range can provide insight regarding the homogeneity of the gas
from which the observed stars formed.  In order to help address these issues
in the bulge, we have measured the [O/Fe], [Si/Fe], and [Ca/Fe] ratios for
stars in all three of our bulge fields.

\begin{figure}
\epsscale{0.75}
\plotone{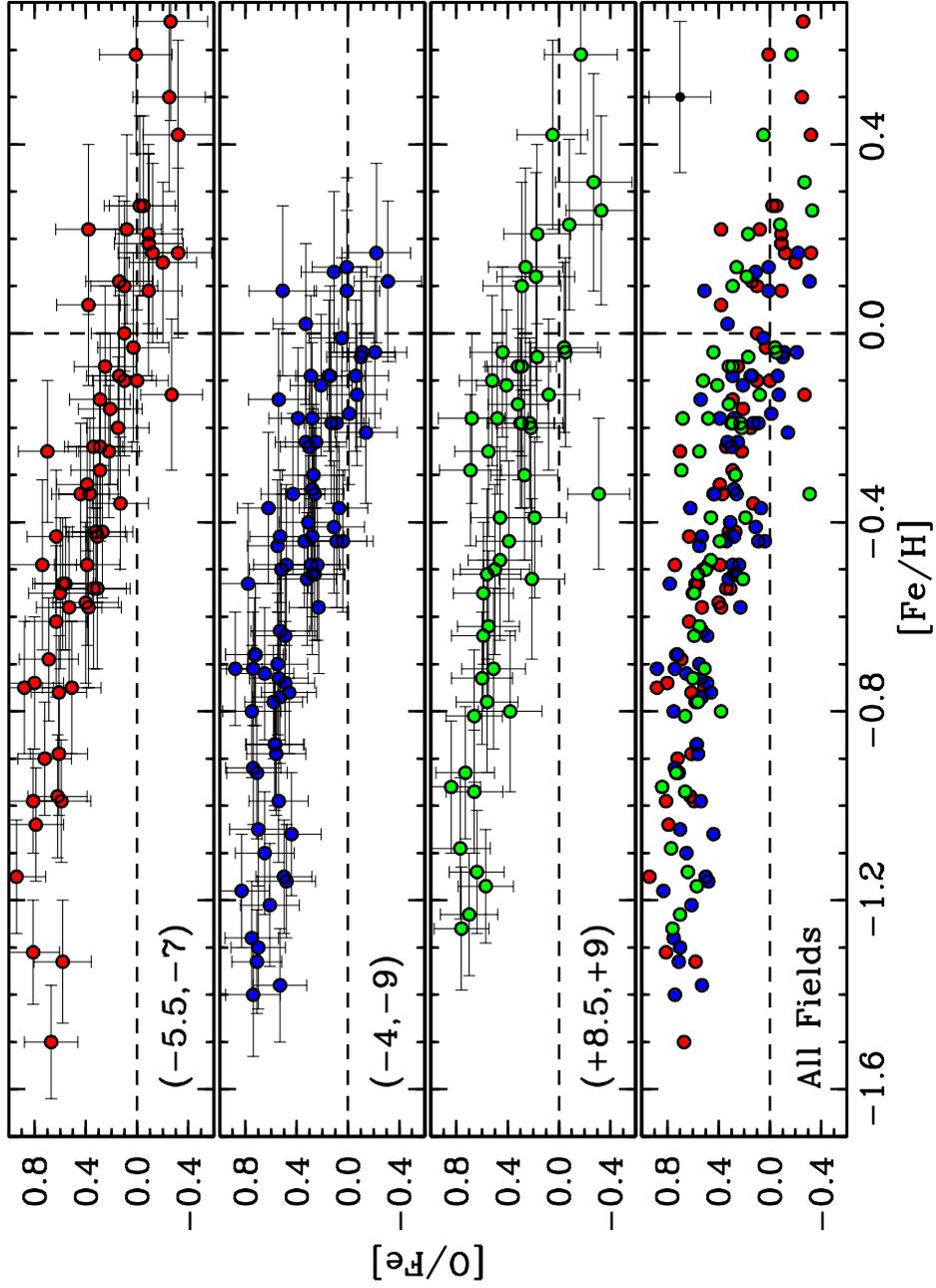}
\caption{[O/Fe] is plotted as a function of [Fe/H] for all three fields.
For display purposes, the individual error bars are suppressed in the bottom
panel and a typical error bar is shown in the top right corner.}
\label{f15}
\end{figure}

\begin{figure}
\epsscale{1.00}
\plotone{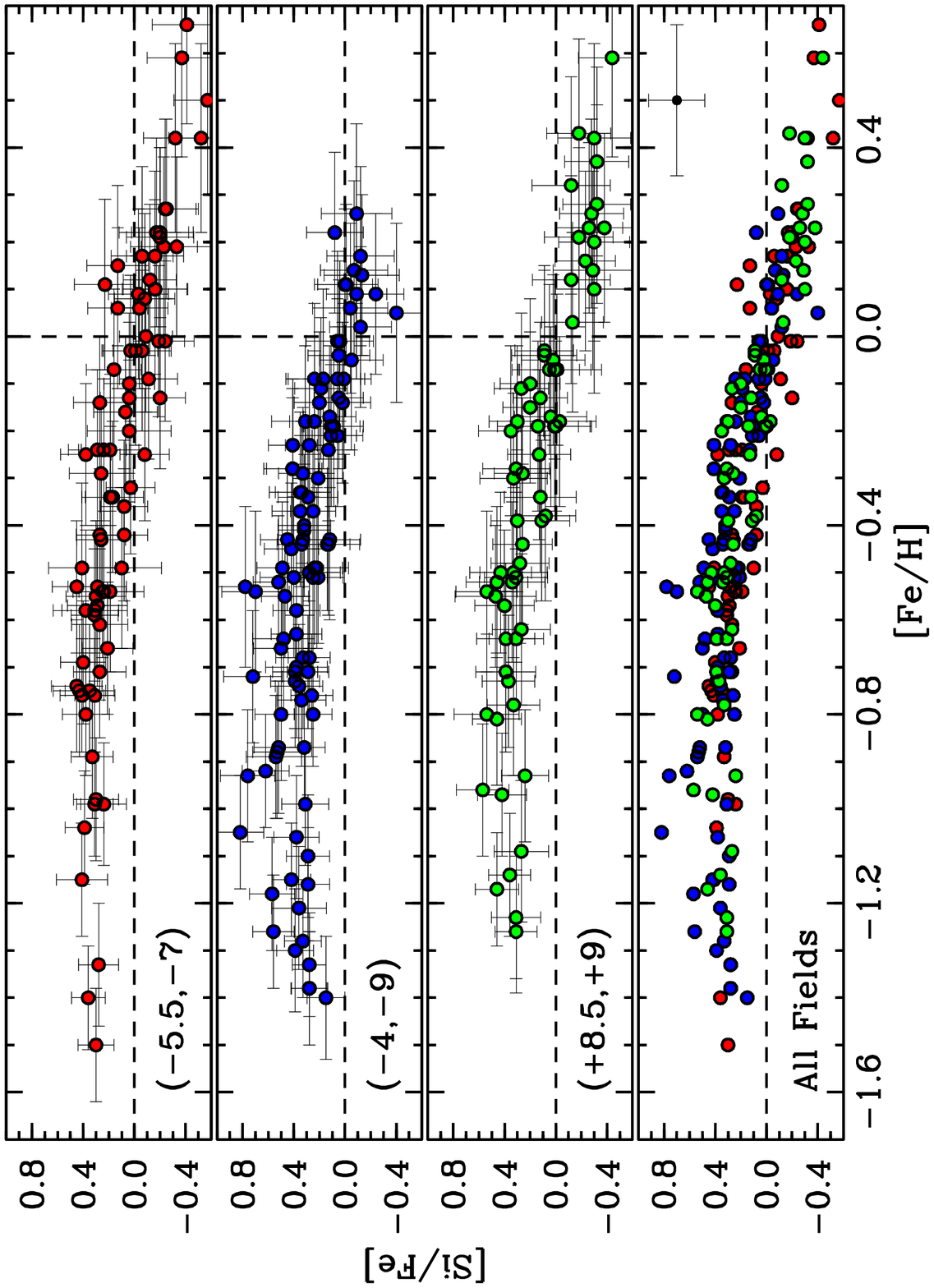}
\caption{Similar to Figure \ref{f15} with [Si/Fe] plotted as a function of 
[Fe/H].}
\label{f16}
\end{figure}

\begin{figure}
\epsscale{1.00}
\plotone{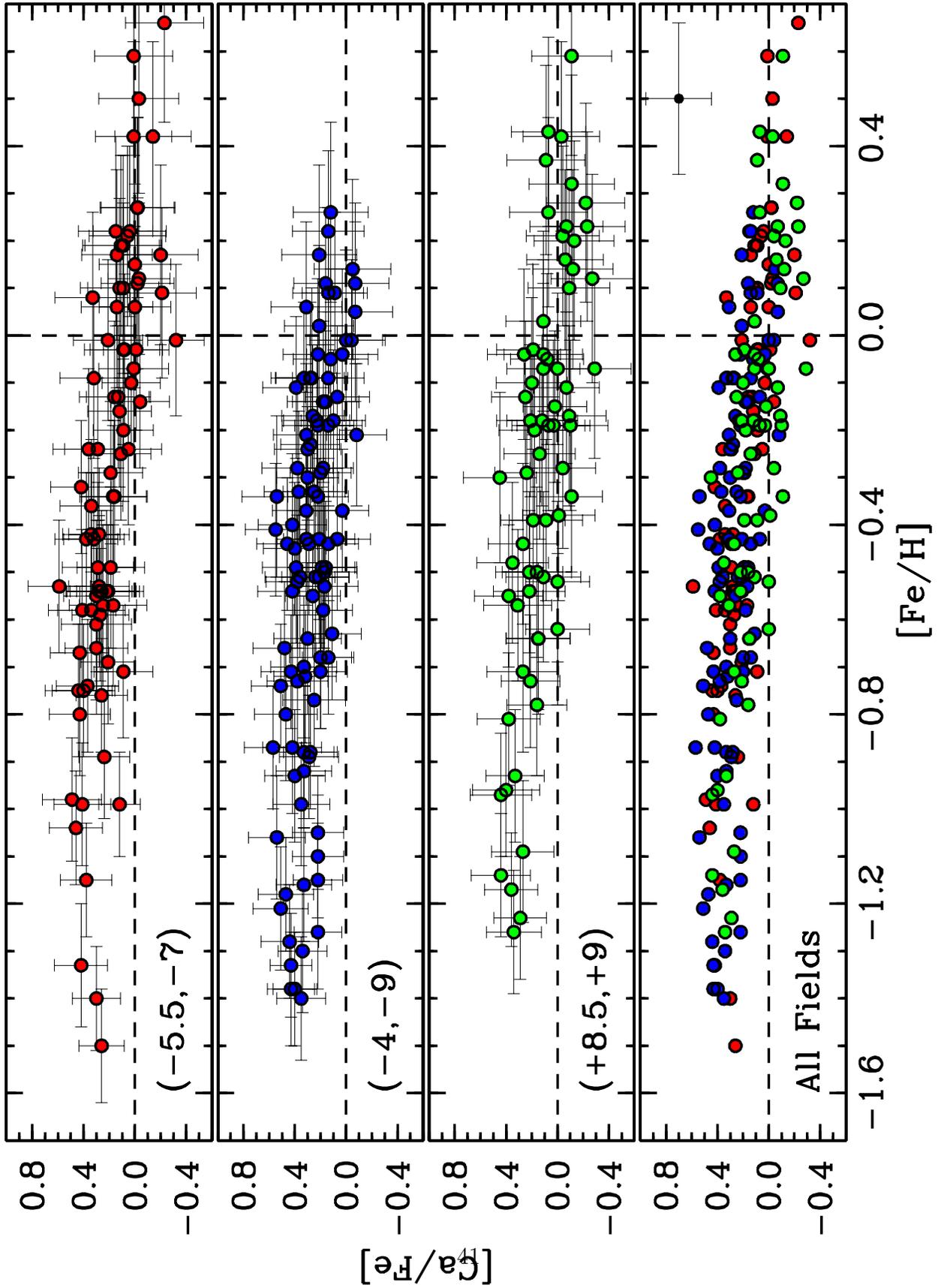}
\caption{Similar to Figure \ref{f15} with [Ca/Fe] plotted as a function of 
[Fe/H].}
\label{f17}
\end{figure}

In Figures \ref{f15}--\ref{f17} we plot [O/Fe], [Si/Fe], and [Ca/Fe] as a 
function of [Fe/H], respectively.  To first order, all of the [X/Fe] ratios
follow the same general trend of being enhanced by $\ga$$+$0.3 dex at 
[Fe/H]$\la$--0.5 and then declining monotonically with increasing [Fe/H].  
However, despite all of the $\alpha$--elements sharing the same [Fe/H] value
at which a steeper decline in their [X/Fe] ratios occurs, it is clear from
Figures \ref{f15}--\ref{f17} that each element exhibits a slightly different
trend.  In particular, we note: (1) the [O/Fe] ratios are $\sim$0.1--0.3 dex 
larger than either [Si/Fe] or [Ca/Fe] for a given [Fe/H], (2) the decline in
[Ca/Fe] with increasing [Fe/H] is more shallow than for [O/Fe] and [Si/Fe],
(3) the [Si/Fe] data appear to continuously decline at [Fe/H]$>$0 while
the [O/Fe] and [Ca/Fe] distributions appear to flatten out, and (4) the 
difference between the median [X/Fe] ratios for stars with [Fe/H]$<$--0.5 and 
[Fe/H]$>$--0.5 is marginally larger for [O/Fe] than either [Si/Fe] or [Ca/Fe]
(see also Table 6).  This last point has been noted previously (e.g., McWilliam
\& Rich 2004; Zoccali et al. 2006; Fulbright et al. 2007; Alves--Brito et al. 
2010; Ryde et al. 2010), and may be related to mass loss in Wolf--Rayet stars, 
rapidly rotating massive metal--poor stars, or metal--poor binary systems 
(e.g., McWilliam et al. 2008; Cescutti et al. 2009; but see also Ryde et al. 
2010 and Alves--Brito et al. 2010).  Unfortunately, we did not measure C, N, 
or Mg in our spectra and cannot comment further on the likelihood of our 
measured [O/Fe] abundances being affected by mass and/or metallicity--dependent
reductions in the oxygen yield from massive stars.

Interestingly, the individual abundance trends are nearly identical in all 
three of our fields, an observation that is highlighted in the bottom panels of 
Figures \ref{f15}--\ref{f17}.  The similarities between the different fields
is also evident by examining Table 6, which summarizes the median [X/Fe]
ratios and star--to--star dispersions for the metal--poor ([Fe/H]$<$--0.5) and
metal--rich ([Fe/H]$>$--0.5) groups in each field.  The dispersion value,
listed as $\sigma$ in Table 6, represents the scatter around a best--fit
line through each subpopulation, and is likely a fair representation of the
measurement error.  On average, the scatter for [O/Fe], [Si/Fe], and [Ca/Fe]
only increases by $+$0.06, $+$0.02, and $+$0.02 dex between the metal--poor
and metal--rich populations.  These data support previous observations 
(e.g., Gonzalez et al. 2011; Johnson et al. 2011; Rich et al. 2012; Uttenthaler
et al. 2012) that the bulge was well--mixed across a large volume and that 
large--scale IMF and star formation history differences were either not present
or were diluted by mixing.

\begin{figure}
\epsscale{1.00}
\plotone{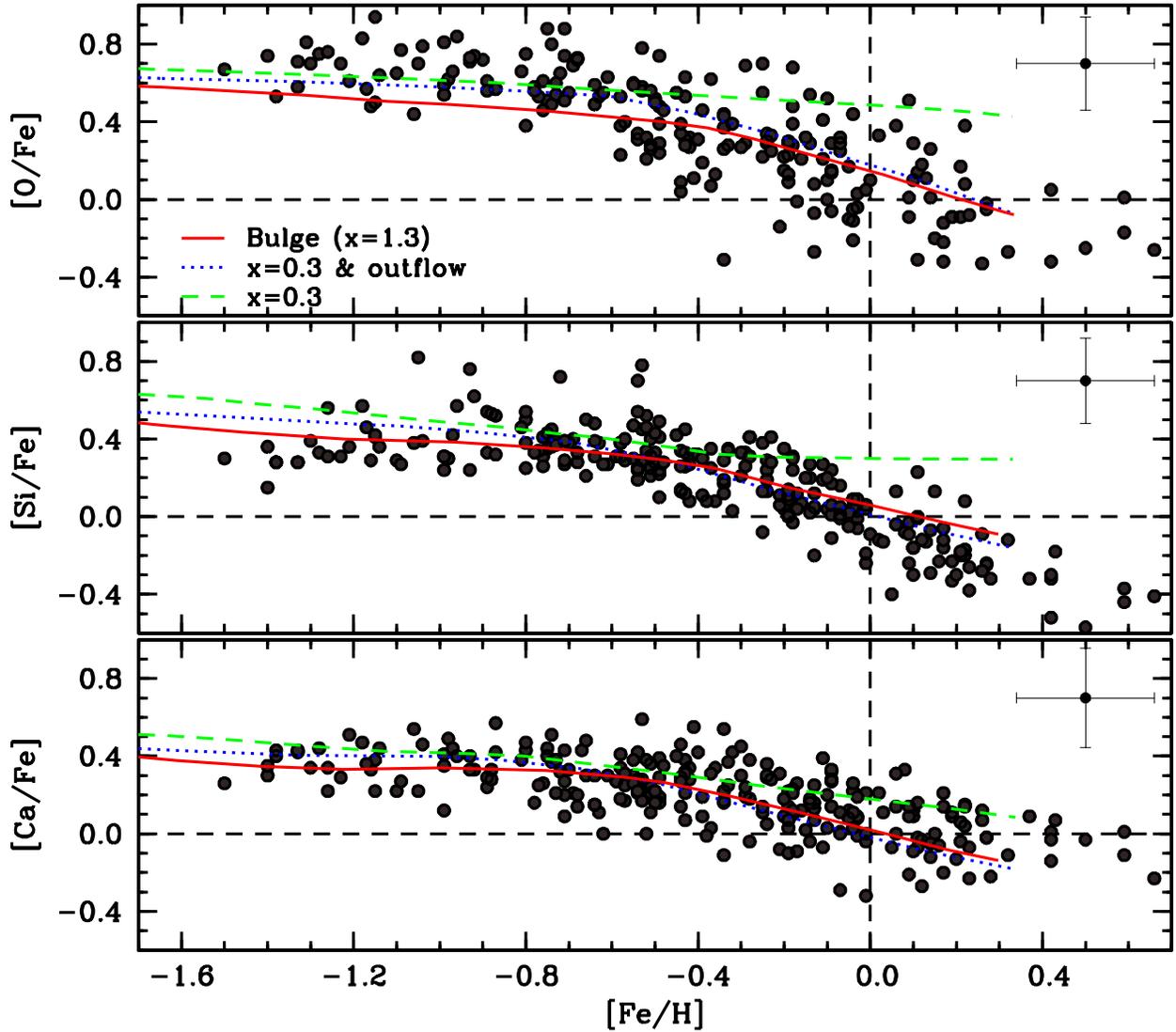}
\caption{The [O/Fe], [Si/Fe], and [Ca/Fe] abundances for all three fields 
(combined) are plotted as a function of [Fe/H].  The solid red line shows the 
predicted change in each $\alpha$ element as a function of [Fe/H], based on the
bulge model of Kobayashi et al. (2011).  The dashed green line is the model
prediction assuming a flatter IMF (x=0.3) and the dotted blue line is the
model prediction assuming x=0.3 with outflow.  Note that the [Si/Fe] model
values have been systematically decreased by 0.2 dex (see text for details).}
\label{f18}
\end{figure}

In order to examine further the small differences between the various 
$\alpha$--element trends, we compare our abundances with three bulge 
chemical enrichment models in Figure \ref{f18}.  The models in Figure \ref{f18}
include the primary bulge model from Kobayashi et al. (2011; solid red line),
a model assuming a flat IMF (x=0.3; dashed green line), and a model assuming a 
flat IMF with both outflow (to reduce the Fe production) and a larger binary 
fraction than the solar neighborhood (dotted blue line).  We find that the 
individual $\alpha$--element abundance trends are fit reasonably well by 
the nominal Kobayashi et al. (2011) model that assumes a Kroupa (2008) IMF 
(x=1.3) for massive stars, rapid star formation (shorter than the solar 
neighborhood), and a short star formation duration ($\sim$3 Gyr).  On the 
other hand, the one--zone flat IMF model clearly fails to reproduce the 
observed decrease in the [X/Fe] ratios with increasing [Fe/H], unless gas 
outflow is included to decrease the Fe yield.  From these data we can conclude 
that: (1) a flat IMF model that does not include outflow (or some other way to 
decrease Fe production) fails to match the universal decline in [$\alpha$/Fe] 
at [Fe/H]$\ga$--0.5 observed in our bulge fields, (2) the data are consistent
with a model in which a significant portion of the bulge formed rapidly ($<$3 
Gyr) and with a majority of its chemical enrichment driven by massive stars, 
and (3) the variations in the [X/Fe] versus [Fe/H] trends between the 
individual $\alpha$--elements are consistent with our current understanding of 
massive star nucleosynthesis.  Although the Kobayashi et al. (2011) model is 
able to provide a reasonable fit to our [O/Fe] and [Ca/Fe] data, we note that 
the model values for [Si/Fe] have been artificially reduced by --0.2 dex 
to provide a consistent fit.  It is not clear if the Si offset is due to 
problems related to the nucleosynthesis codes (e.g., reaction rates; treatments
of convection), the abundance codes (e.g., NLTE/3D effects; model atmosphere 
deficiencies), or both.

\subsubsection{Comparison with Other Bulge Fields}

Several past studies have examined the behavior of various $\alpha$--elements
in fields along both the minor and major axis of the bulge (McWilliam \& Rich 
1994; Rich \& Origlia 2005; Cunha \& Smith 2006; Zoccali et al. 2006; Fulbright
et al. 2007; Lecureur et al. 2007; Rich et al. 2007b; Mel{\'e}ndez et al. 2008; 
Alves--Brito et al. 2010; Bensby et al. 2010a; Ryde et al. 2010; Bensby et al. 
2011; Gonzalez et al. 2011; Hill et al. 2011; Johnson et al. 2011; Bensby et al.
2013; Ness et al. 2013; Rich et al. 2012; Uttenthaler et al. 2012).  All of 
these studies tend to find that metal--poor ([Fe/H]$\la$--0.5) stars 
universally show enhancements of at least $+$0.3 dex in their [$\alpha$/Fe] 
ratios, with the more metal--rich stars exhibiting lower [$\alpha$/Fe] ratios. 
Additionally, despite differing data quality, instruments, and measurement 
techniques, a general consensus has emerged indicating that the [$\alpha$/Fe] 
versus [Fe/H] trends are essentially identical among all bulge fields and that 
this may be especially true among bulge stars with [Fe/H]$\la$--0.5.  

\begin{figure}
\epsscale{1.00}
\plotone{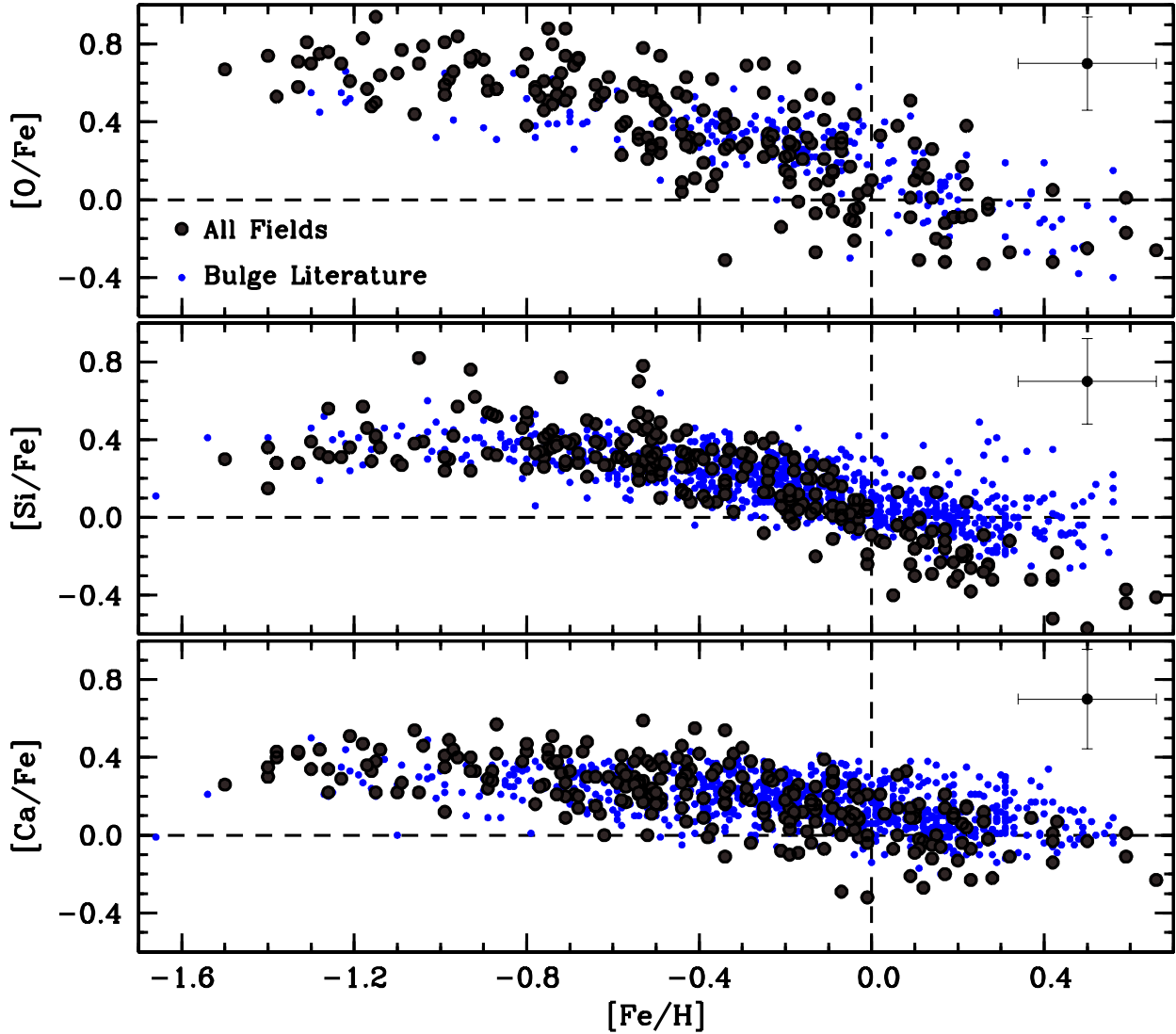}
\caption{[O/Fe], [Si/Fe], and [Ca/Fe] ratios are plotted as a
function of [Fe/H] for all bulge stars measured here (filled grey circles) and
compared with previous bulge measurements (filled blue circles).  The
individual error bars have been suppressed for display purposes and a typical
error bar is shown in the top right corner of each panel.  The literature
data are from: McWilliam \& Rich (1994), Rich \& Origlia (2005), Fulbright et
al. (2007), Lecureur et al. (2007), Rich et al. (2007b), Mel{\'e}ndez et al. 
(2008), Alves--Brito et al. (2010), Bensby et al. (2010a), Ryde et al. (2010), 
Bensby et al. (2011), Gonzalez et al. (2011), Johnson et al. (2011), and Rich 
et al. (2012).}
\label{f19}
\end{figure}

In Figure \ref{f19} we overplot our derived [O/Fe], [Si/Fe], and [Ca/Fe] 
abundances as a function of [Fe/H] with those available in the literature.  
Except for small offsets ($\la$0.1 dex) due to systematic differences in 
abundance scales, our data are generally in good agreement with those available
in other bulge fields.  The lone exception appears to be the [Si/Fe] abundances
at [Fe/H]$\ga$0.  In particular, our [Si/Fe] ratios appear to decline more 
rapidly with increasing [Fe/H] than those found in other fields.  
Unfortunately, the reason for this discrepancy is not immediately clear, but 
it could result from effects such as the high excitation potential of the Si I 
lines used here or the sensitivity of especially the 6155 \AA\ Si I line to CN 
contamination in cool, metal--rich giants (see Figure \ref{f6}).  We note that 
there do not appear to be any strong trends between [Si/Fe] and 
T$_{\rm eff}$ or between abundances derived from the 6145 and 6155 \AA\ Si I 
lines.  Alternatively, the data may reflect a true [Si/Fe] abundance difference
between metal--rich stars on the minor/major bulge axes and in off--axis 
fields.  However, this seems less likely given the agreement between our 
data and the literature for [O/Fe] and [Ca/Fe] at [Fe/H]$>$0.  In any case, the
majority of our data, including the [Si/Fe] ratios at [Fe/H]$<$0, extend the 
findings of previous work (e.g., Gonzalez et al. 2011) indicating that there is
no significant $\alpha$--element gradient present in the bulge.

\subsubsection{Comparison with the Galactic Halo and Disk}

As mentioned in $\S$1, early comparisons of the bulge's chemical composition to 
the other major Galactic stellar populations (i.e., disk and halo) seemed to 
indicate that bulge stars may be uniquely enhanced in their [$\alpha$/Fe] 
ratios (Zoccali et al. 2006; Fulbright et al. 2007; Lecureur et al. 2007), but
recent studies have instead found that the bulge may be chemically similar to 
the thick disk and possibly metal--rich tail of the halo (Prochaska et al. 2000;
Mel{\'e}ndez et al. 2008; Bensby et al. 2010a,b; Ryde et al. 2010; Alves--Brito
et al. 2010; Bensby et al. 2011; Gonzalez et al. 2011; Hill et al. 2011; 
Johnson et al. 2011).  However, it is not yet clear if the chemical similarity,
particularly between the bulge and thick disk, extends to other light and heavy
elements as well (e.g., Fulbright et al. 2007; Lecureur et al. 2007; 
Alves--Brito et al. 2010; Johnson et al. 2012).

\begin{figure}
\epsscale{1.00}
\plotone{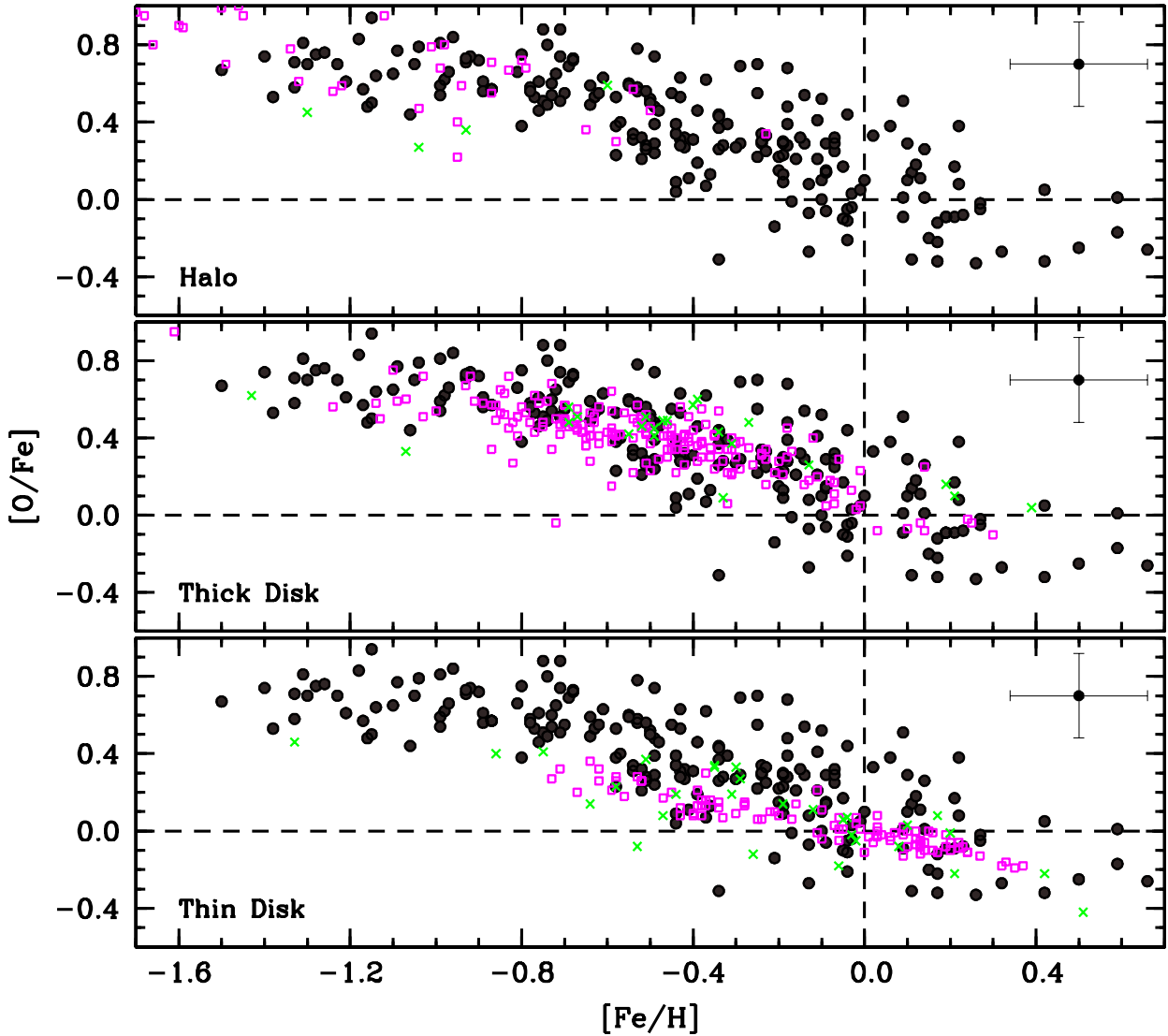}
\caption{[O/Fe] ratios are plotted as a function of [Fe/H] for
bulge stars in all three fields (filled grey circles), the Galactic halo (top 
panels), thick disk (middle panels), and thin disk (bottom panels).  For the 
halo and disk populations, abundances derived from dwarfs and/or subgiants are 
designated by magenta open boxes and abundances derived from giants are 
designated by green crosses.  The individual error bars for our bulge stars 
have been suppressed for display purposes and a typical error bar is shown in 
the top right corner of each panel.  The halo and disk data are from: Tomkin et
al. (1992), Edvardsson et al. (1993), Prochaska et al. (2000), Reddy et al. 
(2003), Bensby et al. (2005), Brewer \& Carney et al. (2006), Reddy et al. 
(2006), and Alves--Brito et al. (2010).  For the Alves--Brito et al. (2010) 
giant data we show only the abundances derived using Kurucz atmospheres.}
\label{f20}
\end{figure}

\begin{figure}
\epsscale{1.00}
\plotone{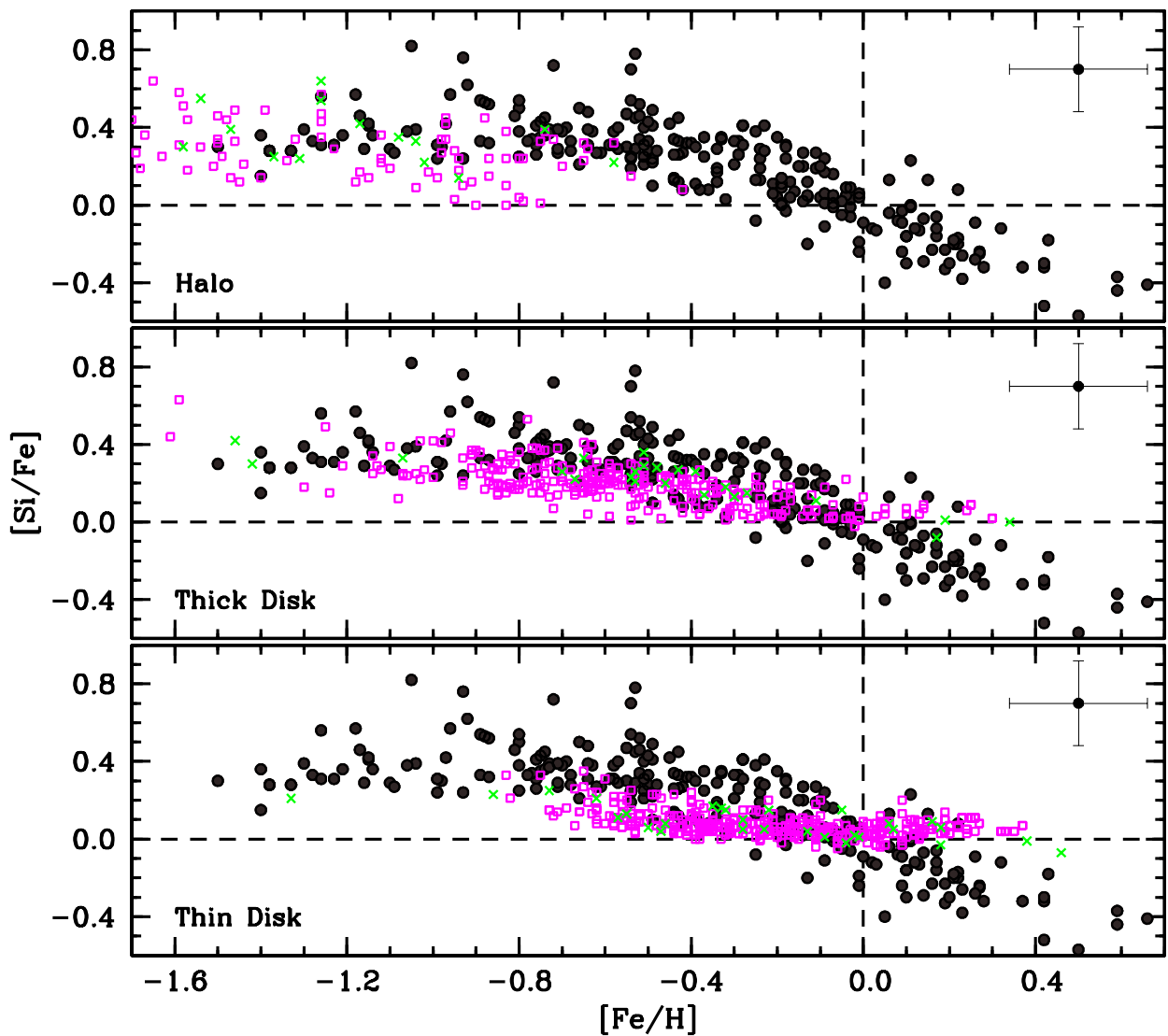}
\caption{Similar to Figure \ref{f20} with [Si/Fe] plotted as a function of 
[Fe/H].  The halo and disk data are from: Edvardsson et al. (1993), Nissen \&
Schuster (1997), Fulbright (2000), Prochaska et al. (2000), Stephens \& 
Boesgaard (2002), Johnson (2002), Bensby et al. (2003), Reddy et al. (2003), 
Bensby et al. (2005), Brewer \& Carney et al. (2006), Reddy et al. (2006), and 
Alves--Brito et al. (2010).}
\label{f21}
\end{figure}

\begin{figure}
\epsscale{1.00}
\plotone{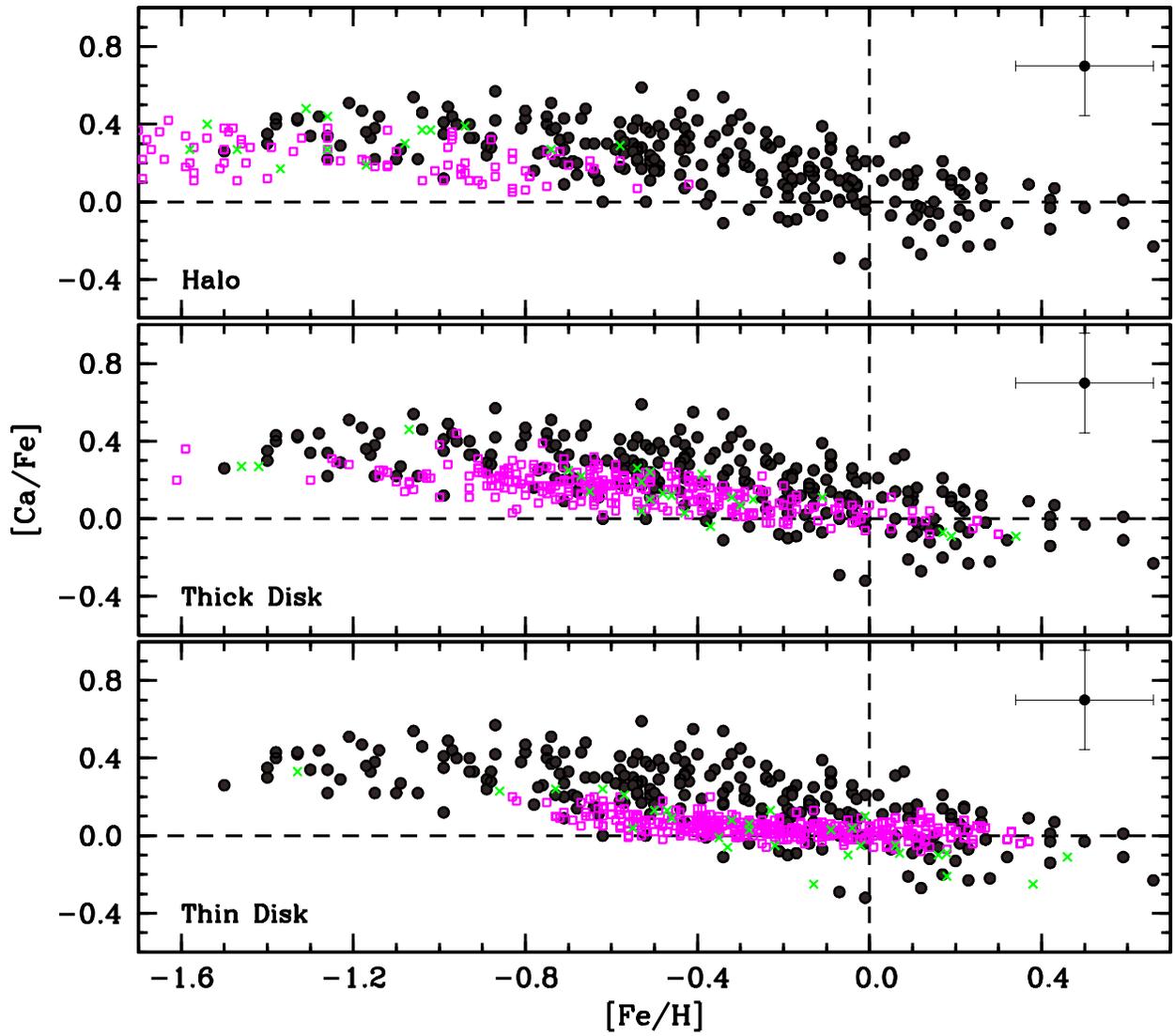}
\caption{Similar to Figure \ref{f20} with [Ca/Fe] plotted as a function of 
[Fe/H].  The literature data are from the same sources as in Figure \ref{f21}.}
\label{f22}
\end{figure}

In Figures \ref{f20}--\ref{f22} we compare our derived [O/Fe], [Si/Fe], and 
[Ca/Fe] abundances to those available in the literature for both giants and 
dwarfs in the halo, thick disk, and thin disk\footnote{As has been noted in 
past work (e.g., Alves--Brito et al. 2010), a direct comparison between 
abundance trends derived from giant and dwarf spectra may be contaminated by
zero--point offsets that can reach at least the 0.1--0.2 dex level.  Therefore,
in Figures \ref{f20}--\ref{f22} we include both giant and dwarf data for 
comparison, but caution that apparent systematic abundance differences 
$\la$0.1 dex in the [$\alpha$/Fe] ratios may not be significant.}.  First 
examining [O/Fe] for stars with [Fe/H]$<$0, Figure \ref{f20} shows that the 
thick disk and halo exhibit nearly identical abundance patterns with respect 
to the bulge, albeit with a smaller star--to--star scatter.  Further inspection
of Figure \ref{f20} also reveals that our most metal--poor bulge stars may have
slightly enhanced ($\sim$0.1--0.2 dex) [O/Fe] ratios relative to similar 
metallicity halo and thick disk giants.  However, we do not believe this offset
is significant because: (1) previous differential analyses have found good 
agreement for [O/Fe] between these populations (e.g., Mel{\'e}ndez et al. 2008;
Alves--Brito et al. 2010), (2) $\sim$0.1 dex of the offset is due to a zero 
point difference between this work and Alves--Brito et al. (2010; the green 
crosses in Figure \ref{f20}), and (3) the observed difference is within the 
error bars of our [O/Fe] measurements.  While the thin disk stars with 
[Fe/H]$\la$0 exhibit lower [O/Fe] ratios than the bulge, the situation becomes 
less clear at higher [Fe/H].  For super--solar metallicities the [O/Fe] 
abundance trends of the thin disk, thick disk, and bulge appear to merge and 
may become indistinguishable.  

For the [Si/Fe] versus [Fe/H] distributions shown in Figure \ref{f21}, we again
find that the halo and thick disk, especially at [Fe/H]$<$0, are nearly 
identical to that of the bulge.  However, much like the [O/Fe] distributions,
any similarities between the bulge and thick disk become difficult to assess
at [Fe/H]$>$0.  The number of thick disk giants in this metallicity regime for
which [Si/Fe] has been measured is too small to draw any firm conclusions.  In 
contrast, the [Si/Fe] ratios of the thin disk stars with [Fe/H]$\la$0 are 
underabundant relative to the bulge, and neither the dwarf nor giant trends 
exhibit the continued decline in [Si/Fe] at [Fe/H]$>$0 seen in both our data 
and possibly those of other bulge studies (see Figure \ref{f19}).

While the halo [Ca/Fe] abundances derived from both dwarfs and giants are 
in excellent agreement with our bulge values, the thick disk [Ca/Fe] ratios 
for stars with [Fe/H]$\la$--0.3 are $\sim$0.1 dex lower.  Similarly, the 
thin disk [Ca/Fe] abundances for stars with [Fe/H]$<$0 are considerably lower 
than those found in the bulge.  However, the [Ca/Fe] abundance differences 
between the thin disk, thick disk, and bulge again become difficult to untangle 
at [Fe/H]$>$0 (see Figure \ref{f22}).  For the more metal--poor thick disk 
giant sample, we can partially reconcile the [Ca/Fe] abundance difference 
by applying the $\sim$0.05 dex zero point offset between our abundance scale 
and that used by Alves--Brito et al. (2010).  The zero point correction also 
affects the thin disk and halo giant samples, but does not alter our 
conclusions regarding the chemical similarities between these populations.

In summary, we find in agreement with recent studies that the metal--poor 
component of the bulge ([Fe/H]$\la$0) and thick disk share indistinguishable
[$\alpha$/Fe] abundance distributions, and that these similarities are 
independent of the $\alpha$--element analyzed.  Additionally, we find that the
most metal--poor ([Fe/H]$<$--1) stars in the bulge also exhibit very similar
[$\alpha$/Fe] distributions to the more metal--rich stars in the halo.  
However, we note that this may not be surprising given that some of our 
metal--poor bulge stars could be halo interlopers (see $\S$4.1.1).  In 
contrast, the thin disk stars are generally underabundant in their 
[$\alpha$/Fe] ratios compared to bulge stars, at least for [Fe/H]$<$0.  For the
most metal--rich stars in the bulge, the [O/Fe] and [Ca/Fe] data suggest that 
there may be significant chemical overlap with the thin disk and/or metal--rich
tail of the thick disk.  While our data and those from past studies indicate 
that the bulge and especially thick disk may have shared similar chemical 
enrichment histories, more work is needed to determine how these two 
populations are linked.

\section{SUMMARY}

We have determined [Fe/H], [O/Fe], [Si/Fe], and [Ca/Fe] abundance ratios, as
well as radial velocities, for 264 RGB stars in three off--axis bulge fields 
centered near (l,b)=(--5.5,--7), (--4,--9), and ($+$8.5,$+$9).  The abundances 
were derived from moderate resolution (R$\approx$18,000), high S/N 
($\sim$75--300 pixel$^{\rm -1}$) spectra obtained with the Hydra multiobject 
spectrographs on the Blanco 4m and WIYN 3.5m telescopes.  While the [Fe/H] 
abundances were measured using equivalent width analyses, the [O/Fe], [Si/Fe], 
and [Ca/Fe] abundances were measured using spectrum synthesis techniques.
The stars were selected from the blue side of the giant branch, but an 
analysis of the color--metallicity relation in other bulge samples that span
a broader color range suggests this may not introduce a strong metallicity
bias.

The metallicity distribution functions for all three fields are in general
agreement with those found in minor--axis, outer bulge fields.  In particular,
the full range in [Fe/H] found in each field spans from roughly [Fe/H]=--1.5 to
$+$0.5 dex, and the median metallicity for the two southern bulge fields
decreases with increasing Galactic latitude.  Combining our (--5.5,--7) and 
(--4,--9) metallicity distributions with the Gonzalez et al. (2011) data at
($+$5.25,--3) indicates that the vertical metallicity gradient observed by
Zoccali et al. (2008) along the minor--axis is also present in off--axis 
fields.  We find a similar [Fe/H] gradient of $\sim$0.4 dex kpc$^{\rm -1}$. 
However, there does not seem to be a similarly strong radial [Fe/H] gradient. 
Interestingly, the northern bulge field at ($+$8.5,$+$9) was found to have the 
highest median metallicity ([Fe/H]=--0.23) sample.  However, it is not clear 
if this is a selection effect resulting from redder stars being observed in 
this field, an indication of asymmetry in the metallicity distribution of the 
bulge, or possible contamination in the field by the X--shaped (and possibly 
metal--rich) bulge structure.  

The radial velocities and dispersion derived here are in good agreement with 
the BRAVA survey, and by extension are in reasonable agreement with the 
Shen et al. (2010) model.  All three of our fields are also found to exhibit
nearly identical radial velocity dispersion versus [Fe/H] relations that show
a strong decrease in dispersion with increasing metallicity.  This trend 
matches previous observations of outer bulge fields, but contrasts with the
dispersion--metallicity relation found at $\mid$b$\mid$$<$6$\degr$.  In the 
two southern bulge fields, we do not find any significant variations with 
respect to the median velocity as a function of [Fe/H].  The same is also true
for the more metal--poor and metal--rich stars in the northern field, but there
is an unexpected increase in median velocity for stars with --0.5$<$[Fe/H]$<$0
in this field.  It is possible that this effect could be tied to the X--shaped 
bulge structure that resides near the ($+$8.5,$+$9) sight line.

We find in agreement with past studies that the [$\alpha$/Fe] ratios are 
enhanced in all bulge fields at [Fe/H]$\la$--0.5, but the $\alpha$--element 
abundances monotonically decline at higher metallicities.  Additionally, 
the [O/Fe], [Si/Fe], and [Ca/Fe] abundance trends exhibit remarkable 
homogeneity across all fields analyzed in this work and previous studies.
A significant [$\alpha$/Fe] abundance gradient does not appear to exist in the
bulge.  Small variations in the slope and magnitude of the individual 
$\alpha$--element trends are consistent with our current understanding of 
massive star nucleosynthesis, and support the idea that the bulge was 
well--mixed and formed more rapidly than the local thin disk.  However, we find
in agreement with several recent studies that for [Fe/H]$\la$0 the bulge 
[$\alpha$/Fe] ratios are indistinguishable from the halo and thick disk.  
Unfortunately, it is not yet clear if the similar [$\alpha$/Fe] trends between 
especially the bulge and thick disk extend to super--solar metallicities.

\acknowledgements

This research has made use of NASA's Astrophysics Data System Bibliographic 
Services.  This material is based upon work supported by the National Science 
Foundation under award No. AST--1003201 to CIJ.  CAP gratefully acknowledges
support from the Daniel Kirkwood Research Fund at Indiana University.  RMR
acknowledges support from NSF grant AST--0709479 and AST--121120995.  AK 
thanks the Deutsche Forschungsgemeinschaft for funding from  Emmy-Noether 
grant Ko 4161/1.

\onecolumn
\clearpage

\clearpage

\clearpage

\clearpage

\clearpage

\clearpage

\clearpage

\clearpage

\clearpage

\clearpage

\clearpage

\clearpage

\clearpage

\clearpage

\clearpage

\clearpage

\clearpage

\clearpage

\clearpage

\clearpage

\clearpage

\clearpage

\input{tab1.tex}
\clearpage
\input{tab2.tex}
\clearpage
\input{tab3.tex}
\clearpage
\input{tab4.tex}
\clearpage
\input{tab5a.tex}
\clearpage
\input{tab5b.tex}
\clearpage
\input{tab5c.tex}
\clearpage
\input{tab5d.tex}
\clearpage
\input{tab6.tex}

\end{document}

%% file: tab1.tex
\tablenum{1}
\tablecolumns{5}
\tablewidth{0pt}

\begin{deluxetable}{ccccc}
\tablecaption{Summary of Hydra Observations}
\tablehead{
\colhead{UT Date}       &
\colhead{Telescope}    &
\colhead{Setup}  &
\colhead{Exposure}        &
\colhead{Field (l,b)}
}

\startdata
2011 Aug 19	&	WIYN	&	6250	&	3x3600 s	&	($+$8.5,$+$9)	\\
2011 Aug 20	&	WIYN	&	6250	&	3x3600 s	&	($+$8.5,$+$9)	\\
2011 Sep 08	&	Blanco	&	6700	&	3x3600 s	&	($-$5.5,$-$7)	\\
2011 Sep 09	&	Blanco	&	6250	&	1x2700 s	&	($-$5.5,$-$7)	\\
2011 Sep 10	&	Blanco	&	6250	&	4x3600 s	&	($-$5.5,$-$7)	\\
2011 Sep 11	&	Blanco	&	6700	&	3x3600 s	&	($-$4,$-$9)	\\
2011 Sep 11	&	Blanco	&	6250	&	1x3600 s	&	($-$4,$-$9)	\\
2011 Sep 12	&	Blanco	&	6250	&	4x3600 s	&	($-$4,$-$9)	\\
\enddata

\end{deluxetable}

%% file: tab2.tex
\tablenum{2}
\tablecolumns{10}
\tablewidth{0pt}

\begin{deluxetable}{cccccccccc}
\tablecaption{Star Identifiers, Photometry, Model Atmosphere Parameters, and 
Radial Velocities}
\rotate
\tablehead{
\colhead{Star ID}	&
\colhead{E(B--V)}      &
\colhead{J}      &
\colhead{H}      &
\colhead{K$_{\rm S}$}      &
\colhead{T$_{\rm eff}$}      &
\colhead{log(g)}      &
\colhead{[Fe/H]}      &
\colhead{V$_{\rm t}$}      &
\colhead{RV Helio.}      \\
\colhead{2MASS}      &
\colhead{}      &
\colhead{}      &
\colhead{}      &
\colhead{}      &
\colhead{(K)}      &
\colhead{(cgs)}      &
\colhead{}      &
\colhead{(km s$^{\rm -1}$)}      &
\colhead{(km s$^{\rm -1}$)}      
}

\startdata
\hline
\multicolumn{10}{c}{(l,b)=(--5.5,--7)}       \\
\hline
 18011989$-$3723462	&	0.212	&	10.996	&	10.313	&	10.148	&	4405	&	1.20	&	$-$1.50	&	1.55	&	$+$50.3	\\
 18003452$-$3729217	&	0.232	&	11.809	&	11.142	&	11.022	&	4605	&	1.60	&	$-$1.40	&	2.15	&	$-$78.2	\\
 18011692$-$3722508	&	0.222	&	11.436	&	10.753	&	10.608	&	4475	&	1.45	&	$-$1.33	&	1.75	&	$+$49.0	\\
 18005584$-$3739103	&	0.212	&	11.525	&	10.805	&	10.659	&	4360	&	1.30	&	$-$1.31	&	1.80	&	$+$171.3	\\
 18015010$-$3735279	&	0.196	&	11.568	&	10.880	&	10.736	&	4425	&	1.50	&	$-$1.15	&	1.80	&	$-$45.9	\\
 18005916$-$3727585	&	0.220	&	11.610	&	10.935	&	10.796	&	4510	&	1.70	&	$-$1.04	&	1.55	&	$+$227.8	\\
 18005027$-$3707505	&	0.268	&	11.908	&	11.261	&	11.110	&	4630	&	1.90	&	$-$0.99	&	1.85	&	$+$48.4	\\
\enddata

\end{deluxetable}

%% file: tab3.tex
\tablenum{3}
\tablecolumns{5}
\tablewidth{0pt}

\begin{deluxetable}{ccccc}
\tablecaption{Chemical Abundance Ratios}
\tablehead{
\colhead{Star ID}	&
\colhead{[Fe/H]}      &
\colhead{[O/Fe]}      &
\colhead{[Si/Fe]}	&
\colhead{[Ca/Fe]}	\\
\colhead{2MASS}      &
\colhead{}      &
\colhead{}      &
\colhead{}      &
\colhead{}      
}

\startdata
\hline
\multicolumn{5}{c}{(l,b)=(--5.5,--7)}       \\
\hline
 18011989$-$3723462	&	$-$1.50	&	$+$0.67	&	$+$0.30	&	$+$0.26	\\
 18003452$-$3729217	&	$-$1.40	&	\nodata	&	$+$0.36	&	$+$0.30	\\
 18011692$-$3722508	&	$-$1.33	&	$+$0.58	&	$+$0.28	&	$+$0.42	\\
 18005584$-$3739103	&	$-$1.31	&	$+$0.81	&	\nodata	&	\nodata	\\
 18015010$-$3735279	&	$-$1.15	&	$+$0.94	&	$+$0.41	&	$+$0.38	\\
 18005916$-$3727585	&	$-$1.04	&	$+$0.79	&	$+$0.39	&	$+$0.46	\\
 18005234$-$3715478	&	$-$0.99	&	$+$0.59	&	$+$0.24	&	$+$0.41	\\
\enddata

\end{deluxetable}

%% file: tab4.tex
\tablenum{4}
\tablecolumns{5}
\tablewidth{0pt}

\begin{deluxetable}{ccccc}
\tablecaption{Line list}
\tablehead{
\colhead{Element}      &
\colhead{Wavelength}      &
\colhead{E.P.}      &
\colhead{log gf}      &
\colhead{Arcturus [Fe/H] or [X/Fe]} \\
\colhead{}      &
\colhead{(\AA)}      &
\colhead{(eV)}      &
\colhead{}      &
\colhead{}      
}

\startdata
Fe I	&	6056.00	&	4.73	&	$-$0.40	&	$-$0.50	\\
Fe I	&	6062.85	&	2.18	&	$-$4.00	&	$-$0.50	\\
Fe I	&	6065.48	&	2.61	&	$-$1.56	&	$-$0.50	\\
Fe I	&	6078.49	&	4.80	&	$-$0.46	&	$-$0.50	\\
Fe I	&	6079.01	&	4.65	&	$-$1.13	&	$-$0.50	\\
Fe I	&	6093.64	&	4.61	&	$-$1.47	&	$-$0.50	\\
Fe I	&	6094.37	&	4.65	&	$-$1.70	&	$-$0.50	\\
Fe I	&	6096.66	&	3.98	&	$-$1.96	&	$-$0.50	\\
\enddata

\end{deluxetable}

%% file: tab5a.tex
\tablenum{5a}
\tablecolumns{7}
\tablewidth{0pt}

\begin{deluxetable}{ccccccc}
\rotate
\tablecaption{Abundance Uncertainty: Iron}
\tablehead{
\colhead{Star ID}	&
\colhead{$\Delta$T$_{\rm eff}$}      &
\colhead{$\Delta$log(g)}      &
\colhead{$\Delta$[M/H]}	&
\colhead{$\Delta$V$_{\rm t}$} &
\colhead{$\sigma$/$\surd$(N)}	&
\colhead{$\Delta$Total}	\\
\colhead{2MASS}      &
\colhead{$+$100 K}      &
\colhead{$+$0.3 cgs}      &
\colhead{$+$0.3 dex}      &
\colhead{$+$0.3 km s$^{\rm -1}$}      &
\colhead{}	&
\colhead{}
}

\startdata
\hline
\multicolumn{7}{c}{(l,b)=(--5.5,--7)}       \\
\hline
 18011989$-$3723462	&	   $+$0.10	&	 $+$0.02	&	  $+$0.01	&	   $-$0.06	&	0.03	&	0.12	\\
 18003452$-$3729217	&	   $+$0.10	&	 $+$0.01	&	  $+$0.00	&	   $-$0.04	&	0.03	&	0.11	\\
 18011692$-$3722508	&	   $+$0.10	&	 $+$0.03	&	  $+$0.02	&	   $-$0.06	&	0.04	&	0.13	\\
 18005584$-$3739103	&	   $+$0.08	&	 $+$0.03	&	  $+$0.03	&	   $-$0.06	&	0.03	&	0.11	\\
 18015010$-$3735279	&	   $+$0.08	&	 $+$0.04	&	  $+$0.03	&	   $-$0.07	&	0.03	&	0.12	\\
 18005916$-$3727585	&	   $+$0.08	&	 $+$0.03	&	  $+$0.03	&	   $-$0.07	&	0.03	&	0.12	\\
 18005027$-$3707505	&	   $+$0.08	&	 $+$0.03	&	  $+$0.02	&	   $-$0.06	&	0.04	&	0.11	\\
\enddata

\end{deluxetable}

%% file: tab5b.tex
\tablenum{5b}
\tablecolumns{7}
\tablewidth{0pt}

\begin{deluxetable}{ccccccc}
\rotate
\tablecaption{Abundance Uncertainty: Oxygen}
\tablehead{
\colhead{Star ID}	&
\colhead{$\Delta$T$_{\rm eff}$}      &
\colhead{$\Delta$log(g)}      &
\colhead{$\Delta$[M/H]}	&
\colhead{$\Delta$V$_{\rm t}$} &
\colhead{$\sigma$/$\surd$(N)}	&
\colhead{$\Delta$Total}	\\
\colhead{2MASS}      &
\colhead{$+$100 K}      &
\colhead{$+$0.3 cgs}      &
\colhead{$+$0.3 dex}      &
\colhead{$+$0.3 km s$^{\rm -1}$}      &
\colhead{}	&
\colhead{}
}

\startdata
\hline
\multicolumn{7}{c}{(l,b)=(--5.5,--7)}       \\
\hline
 18011989$-$3723462	&	   $+$0.01	&	 $+$0.12	&	 $+$0.11	&	  $-$0.01	&	0.05	&	0.17	\\
 18003452$-$3729217	&	\nodata	&	\nodata	&	\nodata	&	\nodata	&	\nodata	&	\nodata	\\
 18011692$-$3722508	&	   $+$0.02	&	 $+$0.13	&	 $+$0.11	&	  $-$0.01	&	0.05	&	0.18	\\
 18005584$-$3739103	&	   $+$0.01	&	 $+$0.12	&	 $+$0.11	&	  $-$0.02	&	0.05	&	0.17	\\
 18015010$-$3735279	&	   $+$0.01	&	 $+$0.14	&	 $+$0.11	&	  $-$0.03	&	0.05	&	0.19	\\
 18005916$-$3727585	&	   $+$0.02	&	 $+$0.12	&	 $+$0.12	&	  $-$0.02	&	0.05	&	0.18	\\
 18005027$-$3707505	&	   $+$0.01	&	 $+$0.13	&	 $+$0.11	&	  $-$0.02	&	0.05	&	0.18	\\
\enddata

\end{deluxetable}

%% file: tab5c.tex
\tablenum{5c}
\tablecolumns{7}
\tablewidth{0pt}

\begin{deluxetable}{ccccccc}
\rotate
\tablecaption{Abundance Uncertainty: Silicon}
\tablehead{
\colhead{Star ID}	&
\colhead{$\Delta$T$_{\rm eff}$}      &
\colhead{$\Delta$log(g)}      &
\colhead{$\Delta$[M/H]}	&
\colhead{$\Delta$V$_{\rm t}$} &
\colhead{$\sigma$/$\surd$(N)}	&
\colhead{$\Delta$Total}	\\
\colhead{2MASS}      &
\colhead{$+$100 K}      &
\colhead{$+$0.3 cgs}      &
\colhead{$+$0.3 dex}      &
\colhead{$+$0.3 km s$^{\rm -1}$}      &
\colhead{}	&
\colhead{}
}

\startdata
\hline
\multicolumn{7}{c}{(l,b)=(--5.5,--7)}       \\
\hline
 18011989$-$3723462	&	   $-$0.02	&	  $+$0.03	&	  $+$0.04	&	  $-$0.01	&	0.05	&	0.07	\\
 18003452$-$3729217	&	  $+$0.00	&	  $+$0.04	&	  $+$0.03	&	  $-$0.01	&	0.05	&	0.07	\\
 18011692$-$3722508	&	  $+$0.00	&	  $+$0.06	&	  $+$0.04	&	  $-$0.01	&	0.05	&	0.09	\\
 18005584$-$3739103	&	\nodata	&	\nodata	&	\nodata	&	\nodata	&	\nodata	&	\nodata	\\
 18015010$-$3735279	&	   $-$0.03	&	  $+$0.08	&	  $+$0.05	&	  $-$0.02	&	0.12	&	0.16	\\
 18005916$-$3727585	&	   $-$0.02	&	  $+$0.04	&	  $+$0.06	&	  $-$0.02	&	0.05	&	0.09	\\
 18005027$-$3707505	&	   $-$0.02	&	  $+$0.06	&	  $+$0.05	&	  $-$0.02	&	0.03	&	0.09	\\
\enddata

\end{deluxetable}

%% file: tab5d.tex
\tablenum{5d}
\tablecolumns{7}
\tablewidth{0pt}

\begin{deluxetable}{ccccccc}
\rotate
\tablecaption{Abundance Uncertainty: Calcium}
\tablehead{
\colhead{Star ID}	&
\colhead{$\Delta$T$_{\rm eff}$}      &
\colhead{$\Delta$log(g)}      &
\colhead{$\Delta$[M/H]}	&
\colhead{$\Delta$V$_{\rm t}$} &
\colhead{$\sigma$/$\surd$(N)}	&
\colhead{$\Delta$Total}	\\
\colhead{2MASS}      &
\colhead{$+$100 K}      &
\colhead{$+$0.3 cgs}      &
\colhead{$+$0.3 dex}      &
\colhead{$+$0.3 km s$^{\rm -1}$}      &
\colhead{}	&
\colhead{}
}

\startdata
\hline
\multicolumn{7}{c}{(l,b)=(--5.5,--7)}       \\
\hline
 18011989$-$3723462	&	  $+$0.09	&	  $-$0.02	&	   $-$0.03	&	 $-$0.07	&	0.06	&	0.13	\\
 18003452$-$3729217	&	  $+$0.09	&	  $-$0.01	&	   $-$0.02	&	 $-$0.05	&	0.10	&	0.15	\\
 18011692$-$3722508	&	  $+$0.11	&	  $-$0.01	&	   $-$0.02	&	 $-$0.09	&	0.06	&	0.16	\\
 18005584$-$3739103	&	\nodata	&	\nodata	&	\nodata	&	\nodata	&	\nodata	&	\nodata	\\
 18015010$-$3735279	&	  $+$0.10	&	  $-$0.01	&	   $-$0.02	&	 $-$0.11	&	0.04	&	0.16	\\
 18005916$-$3727585	&	  $+$0.11	&	  $-$0.02	&	   $-$0.02	&	 $-$0.12	&	0.03	&	0.17	\\
 18005027$-$3707505	&	  $+$0.09	&	  $-$0.01	&	   $-$0.02	&	 $-$0.07	&	0.03	&	0.12	\\
\enddata

\end{deluxetable}

%% file: tab6.tex
\tablenum{6}
\tablecolumns{7}
\tablewidth{0pt}

\begin{deluxetable}{ccccccc}
\tablecaption{$\alpha$--Element Abundance Summary}
\tablehead{
\colhead{Field}	&
\colhead{[O/Fe]}      &
\colhead{$\sigma$}      &
\colhead{[Si/Fe]}      &
\colhead{$\sigma$}      &
\colhead{[Ca/Fe]}      &
\colhead{$\sigma$}	\\
\colhead{} &
\colhead{Median} &
\colhead{} &
\colhead{Median} &
\colhead{} &
\colhead{Median} &
\colhead{}
}

\startdata
\hline
\multicolumn{7}{c}{[Fe/H]$<$--0.5}       \\
\hline
($-$5.5,$-$7)	&	$+$0.61	&	0.14	&	$+$0.31	&	0.07	&	$+$0.30	&	0.11	\\
($-$4,$-$9)	&	$+$0.57	&	0.14	&	$+$0.38	&	0.16	&	$+$0.33	&	0.11	\\
($+$8.5,$+$9)	&	$+$0.59	&	0.11	&	$+$0.38	&	0.09	&	$+$0.27	&	0.10	\\
\hline
\multicolumn{7}{c}{[Fe/H]$>$--0.5}       \\
\hline
($-$5.5,$-$7)	&	$+$0.14	&	0.18	&	$-$0.03	&	0.13	&	$+$0.09	&	0.12	\\
($-$4,$-$9)	&	$+$0.22	&	0.19	&	$+$0.13	&	0.12	&	$+$0.21	&	0.13	\\
($+$8.5,$+$9)	&	$+$0.27	&	0.22	&	$+$0.02	&	0.11	&	$+$0.05	&	0.14	\\
\enddata

\tablecomments{The $\sigma$ values listed in the table represent the scatter
around a best fit line to either the [Fe/H]$<$--0.5 or [Fe/H]$>$--0.5 
stars in each field.}
\end{deluxetable}

%% file: ms.bbl
\begin{thebibliography}{}

\bibitem[Alonso et al.(1999)]{1999A&AS..140..261A} Alonso, A., Arribas, S., \& Mart{\'{\i}}nez-Roger, C.\ 1999, \aaps, 140, 261 

\bibitem[Alonso et al.(2001)]{2001A&A...376.1039A} Alonso, A., Arribas, S., \& Mart{\'{\i}}nez-Roger, C.\ 2001, \aap, 376, 1039

\bibitem[Alves-Brito et al.(2010)]{2010A&A...513A..35A} Alves-Brito, A., Mel{\'e}ndez, J., Asplund, M., Ram{\'{\i}}rez, I., \& Yong, D.\ 2010, \aap, 513, A35 

\bibitem[Babusiaux et al.(2010)]{2010A&A...519A..77B} Babusiaux, C., G{\'o}mez, A., Hill, V., et al.\ 2010, \aap, 519, A77

\bibitem[Bensby et al.(2003)]{2003A&A...410..527B} Bensby, T., Feltzing, S.,
\& Lundstr{\"o}m, I.\ 2003, \aap, 410, 527

\bibitem[Bensby et al.(2005)]{2005A&A...433..185B} Bensby, T., Feltzing, S., Lundstr{\"o}m, I., \& Ilyin, I.\ 2005, \aap, 433, 185

\bibitem[Bensby et al.(2010)]{2010A&A...512A..41B} Bensby, T., Feltzing, S., Johnson, J.~A., et al.\ 2010a, \aap, 512, A41

\bibitem[Bensby et al.(2010)]{2010A&A...516L..13B} Bensby, T., Alves-Brito, A., Oey, M.~S., Yong, D., \& Mel{\'e}ndez, J.\ 2010b, \aap, 516, L13

\bibitem[Bensby et al.(2011)]{2011A&A...533A.134B} Bensby, T., Ad{\'e}n, D., Mel{\'e}ndez, J., et al.\ 2011, \aap, 533, A134

\bibitem[Bensby et al.(2013)]{2013A&A...549A.147B} Bensby, T., Yee, J.~C., Feltzing, S., et al.\ 2013, \aap, 549, A147

\bibitem[Bershady et al.(2008)]{2008SPIE.7014E..15B} Bershady, M., Barden, 
S., Blanche, P.-A., et al.\ 2008, \procspie, 7014, 15

\bibitem[Brewer \& Carney(2006)]{2006AJ....131..431B} Brewer, M.-M., \& Carney, B.~W.\ 2006, \aj, 131, 431

\bibitem[Castelli et al.(1997)]{1997A&A...318..841C} Castelli, F., Gratton, R.~G., \& Kurucz, R.~L.\ 1997, \aap, 318, 841

\bibitem[Cescutti et al.(2009)]{2009A&A...505..605C} Cescutti, G., Matteucci, F., McWilliam, A., \& Chiappini, C.\ 2009, \aap, 505, 605

\bibitem[Clarkson et al.(2008)]{2008ApJ...684.1110C} Clarkson, W., Sahu, 
K., Anderson, J., et al.\ 2008, \apj, 684, 1110 

\bibitem[Cunha \& Smith(2006)]{2006ApJ...651..491C} Cunha, K., \& Smith, V.~V.\ 2006, \apj, 651, 491

\bibitem[De Propris et al.(2011)]{2011ApJ...732L..36D} De Propris, R., 
Rich, R.~M., Kunder, A., et al.\ 2011, \apjl, 732, L36 

\bibitem[Dotter et al.(2007)]{2007AJ....134..376D} Dotter, A., Chaboyer, 
B., Jevremovi{\'c}, D., et al.\ 2007, \aj, 134, 376

\bibitem[Edvardsson et al.(1993)]{1993A&A...275..101E} Edvardsson, B., Andersen, J., Gustafsson, B., et al.\ 1993, \aap, 275, 101

\bibitem[Fiorucci \& Munari(2003)]{2003A&A...401..781F} Fiorucci, M., \& Munari, U.\ 2003, \aap, 401, 781

\bibitem[Fulbright(2000)]{2000AJ....120.1841F} Fulbright, J.~P.\ 2000, \aj, 
120, 1841

\bibitem[Fulbright et al.(2006)]{2006ApJ...636..821F} Fulbright, J.~P., 
McWilliam, A., \& Rich, R.~M.\ 2006, \apj, 636, 821

\bibitem[Fulbright et al.(2007)]{2007ApJ...661.1152F} Fulbright, J.~P., 
McWilliam, A., \& Rich, R.~M.\ 2007, \apj, 661, 1152

\bibitem[Girard et al.(2011)]{2011AJ....142...15G} Girard, T.~M., van 
Altena, W.~F., Zacharias, N., et al.\ 2011, \aj, 142, 15

\bibitem[Gonz{\'a}lez Hern{\'a}ndez \& Bonifacio(2009)]{2009A&A...497..497G} Gonz{\'a}lez Hern{\'a}ndez, J.~I., \& Bonifacio, P.\ 2009, \aap, 497, 497 

\bibitem[Gonzalez et al.(2011)]{2011A&A...530A..54G} Gonzalez, O.~A., Rejkuba, M., Zoccali, M., et al.\ 2011, \aap, 530, A54

\bibitem[Hinkle et al.(2000)]{2000vnia.book.....H} Hinkle, K., Wallace, L.,
Valenti, J., \& Harmer, D.\ 2000, Visible and Near Infrared Atlas of the Arcturus Spectrum 3727-9300 A ed.~Kenneth Hinkle, Lloyd Wallace, Jeff Valenti, and Dianne Harmer.~(San Francisco: ASP) ISBN: 1-58381-037-4, 2000

\bibitem[Howard et al.(2008)]{2008ApJ...688.1060H} Howard, C.~D., Rich, 
R.~M., Reitzel, D.~B., et al.\ 2008, \apj, 688, 1060 

\bibitem[Johnson(2002)]{2002ApJS..139..219J} Johnson, J.~A.\ 2002, \apjs, 
139, 219

\bibitem[Johnson et al.(2008)]{2008ApJ...681.1505J} Johnson, C.~I., 
Pilachowski, C.~A., Simmerer, J., \& Schwenk, D.\ 2008, \apj, 681, 1505

\bibitem[Johnson et al.(2011)]{2011ApJ...732..108J} Johnson, C.~I., Rich, 
R.~M., Fulbright, J.~P., Valenti, E., \& McWilliam, A.\ 2011, \apj, 732, 108

\bibitem[Johnson et al.(2012)]{2012ApJ...749..175J} Johnson, C.~I., Rich, 
R.~M., Kobayashi, C., \& Fulbright, J.~P.\ 2012, \apj, 749, 175 

\bibitem[Knezek et al.(2010)]{2010SPIE.7735E.240K} Knezek, P.~M., Bershady, 
M.~A., Willmarth, D., et al.\ 2010, \procspie, 7735, 240

\bibitem[Kobayashi et al.(2011)]{2011MNRAS.414.3231K} Kobayashi, C., 
Karakas, A.~I., \& Umeda, H.\ 2011, \mnras, 414, 3231

\bibitem[Kroupa(2008)]{2008ASPC..390....3K} Kroupa, P.\ 2008, Pathways 
Through an Eclectic Universe, 390, 3

\bibitem[Kunder et al.(2012)]{2012AJ....143...57K} Kunder, A., Koch, A., 
Rich, R.~M., et al.\ 2012, \aj, 143, 57

\bibitem[Ku{\v c}inskas et al.(2005)]{2005A&A...442..281K} Ku{\v c}inskas, A., Hauschildt, P.~H., Ludwig, H.-G., et al.\ 2005, \aap, 442, 281 

\bibitem[Ku{\v c}inskas et al.(2006)]{2006A&A...452.1021K} Ku{\v c}inskas, A., Hauschildt, P.~H., Brott, I., et al.\ 2006, \aap, 452, 1021 

\bibitem[Kupka et al.(2000)]{2000BaltA...9..590K} Kupka, F.~G., Ryabchikova, 
T.~A., Piskunov, N.~E., Stempels, H.~C., \& Weiss, W.~W.\ 2000, Baltic 
Astronomy, 9, 590 

\bibitem[Lecureur et al.(2007)]{2007A&A...465..799L} Lecureur, A., Hill, V., Zoccali, M., et al.\ 2007, \aap, 465, 799 

\bibitem[Matteucci \& Brocato(1990)]{1990ApJ...365..539M} Matteucci, F., \& Brocato, E.\ 1990, \apj, 365, 539

\bibitem[McWilliam(1997)]{1997ARA&A..35..503M} McWilliam, A.\ 1997, \araa, 35, 503

\bibitem[McWilliam \& Rich(1994)]{1994ApJS...91..749M} McWilliam, A., \& Rich, R.~M.\ 1994, \apjs, 91, 749

\bibitem[McWilliam \& Rich(2004)]{2004oee..sympE..38M} McWilliam, A., \& Rich, R.~M.\ 2004, Origin and Evolution of the Elements, ed. A. McWilliam \& M. Rausch
(Pasadena, CA: Carnegie Obs.)

\bibitem[McWilliam \& Zoccali(2010)]{2010ApJ...724.1491M} McWilliam, A., \& Zoccali, M.\ 2010, \apj, 724, 1491

\bibitem[McWilliam et al.(2008)]{2008AJ....136..367M} McWilliam, A., 
Matteucci, F., Ballero, S., et al.\ 2008, \aj, 136, 367

\bibitem[McWilliam et al.(2010)]{2010IAUS..265..279M} McWilliam, A., 
Fulbright, J., \& Rich, R.~M.\ 2010, IAU Symposium, 265, 279

\bibitem[Mel{\'e}ndez et al.(2008)]{2008A&A...484L..21M} Mel{\'e}ndez, J., Asplund, M., Alves-Brito, A., et al.\ 2008, \aap, 484, L21

\bibitem[Minniti(1996)]{1996ApJ...459..579M} Minniti, D.\ 1996, \apj, 459, 
579

\bibitem[Moultaka et al.(2004)]{2004PASP..116..693M} Moultaka, J., 
Ilovaisky, S.~A., Prugniel, P., \& Soubiran, C.\ 2004, \pasp, 116, 69

\bibitem[Ness et al.(2012)]{2012ApJ...756...22N} Ness, M., Freeman, K., 
Athanassoula, E., et al.\ 2012, \apj, 756, 22

\bibitem[Ness et al.(2013)]{2013MNRAS.tmp..635N} Ness, M., Freeman, K.,
Athanassoula, E., et al.\ 2013, \mnras, 635

\bibitem[Nissen \& Schuster(1997)]{1997A&A...326..751N} Nissen, P.~E., \& Schuster, W.~J.\ 1997, \aap, 326, 75

\bibitem[Peterson et al.(1993)]{1993ApJ...404..333P} Peterson, R.~C., Dalle 
Ore, C.~M., \& Kurucz, R.~L.\ 1993, \apj, 404, 333

\bibitem[Press et al.(1992)]{1992nrfa.book.....P} Press, W.~H., Teukolsky,
S.~A., Vetterling, W.~T., \& Flannery, B.~P.\ 1992, Cambridge: University Press, |c1992, 2nd ed.

\bibitem[Prochaska et al.(2000)]{2000AJ....120.2513P} Prochaska, J.~X., 
Naumov, S.~O., Carney, B.~W., McWilliam, A., \& Wolfe, A.~M.\ 2000, \aj, 120, 2513

\bibitem[Ralchenko et al.(2011)]{}Ralchenko, Yu., Kramida, A.E., Reader, J.,
and NIST ASD Team (2011).\emph{NIST Atomic Spectra Database} (version 4.1).
National Institute of Standards and Technology, Gaithersburg, MD.

\bibitem[Ram{\'{\i}}rez \& Allende Prieto(2011)]{2011ApJ...743..135R} Ram{\'{\i}}rez, I., \& Allende Prieto, C.\ 2011, \apj, 743, 135 

\bibitem[Reddy et al.(2003)]{2003MNRAS.340..304R} Reddy, B.~E., Tomkin, J., 
Lambert, D.~L., \& Allende Prieto, C.\ 2003, \mnras, 340, 304

\bibitem[Reddy et al.(2006)]{2006MNRAS.367.1329R} Reddy, B.~E., Lambert, 
D.~L., \& Allende Prieto, C.\ 2006, \mnras, 367, 1329

\bibitem[Rich(1990)]{1990ApJ...362..604R} Rich, R.~M.\ 1990, \apj, 362, 604

\bibitem[Rich \& Origlia(2005)]{2005ApJ...634.1293R} Rich, R.~M., \& Origlia, L.\ 2005, \apj, 634, 1293 

\bibitem[Rich et al.(2007)]{2007ApJ...658L..29R} Rich, R.~M., Reitzel, 
D.~B., Howard, C.~D., \& Zhao, H.\ 2007a, \apjl, 658, L29

\bibitem[Rich et al.(2007)]{2007ApJ...665L.119R} Rich, R.~M., Origlia, L., 
\& Valenti, E.\ 2007b, \apjl, 665, L119 

\bibitem[Rich et al.(2012)]{2012ApJ...746...59R} Rich, R.~M., Origlia, L., 
\& Valenti, E.\ 2012, \apj, 746, 59

\bibitem[Robin et al.(2003)]{2003A&A...409..523R} Robin, A.~C., Reyl{\'e}, C., Derri{\`e}re, S., \& Picaud, S.\ 2003, \aap, 409, 523

\bibitem[Ryde et al.(2010)]{2010A&A...509A..20R} Ryde, N., Gustafsson, B., Edvardsson, B., et al.\ 2010, \aap, 509, A20

\bibitem[Saha et al.(2012)]{2012MNRAS.421..333S} Saha, K., 
Martinez-Valpuesta, I., \& Gerhard, O.\ 2012, \mnras, 421, 333

\bibitem[Saito et al.(2011)]{2011AJ....142...76S} Saito, R.~K., Zoccali, 
M., McWilliam, A., et al.\ 2011, \aj, 142, 76

\bibitem[Schlegel et al.(1998)]{1998ApJ...500..525S} Schlegel, D.~J., 
Finkbeiner, D.~P., \& Davis, M.\ 1998, \apj, 500, 525 

\bibitem[Shen et al.(2010)]{2010ApJ...720L..72S} Shen, J., Rich, R.~M., 
Kormendy, J., et al.\ 2010, \apjl, 720, L72

\bibitem[Skrutskie et al.(2006)]{2006AJ....131.1163S} Skrutskie, M.~F., 
Cutri, R.~M., Stiening, R., et al.\ 2006, \aj, 131, 1163 

\bibitem[Sneden(1973)]{1973ApJ...184..839S} Sneden, C.\ 1973, \apj, 184, 839

\bibitem[Soto et al.(2007)]{2007ApJ...665L..31S} Soto, M., Rich, R.~M., 
\& Kuijken, K.\ 2007, \apjl, 665, L31

\bibitem[Stephens \& Boesgaard(2002)]{2002AJ....123.1647S} Stephens, A., \& Boesgaard, A.~M.\ 2002, \aj, 123, 1647

\bibitem[Thevenin(1990)]{1990A&AS...82..179T} Thevenin, F.\ 1990, \aaps, 82, 179

\bibitem[Tinsley(1979)]{1979ApJ...229.1046T} Tinsley, B.~M.\ 1979, \apj, 
229, 1046

\bibitem[Tomkin et al.(1992)]{1992AJ....104.1568T} Tomkin, J., Lemke, M., 
Lambert, D.~L., \& Sneden, C.\ 1992, \aj, 104, 1568

\bibitem[Uttenthaler et al.(2012)]{2012A&A...546A..57U} Uttenthaler, S., Schultheis, M., Nataf, D.~M., et al.\ 2012, \aap, 546, A57

\bibitem[Zoccali et al.(2006)]{2006A&A...457L...1Z} Zoccali, M., Lecureur, A., Barbuy, B., et al.\ 2006, \aap, 457, L1

\bibitem[Zoccali et al.(2008)]{2008A&A...486..177Z} Zoccali, M., Hill, V., Lecureur, A., et al.\ 2008, \aap, 486, 177

\end{thebibliography}
